\documentclass[titlepage,12pt]{utarticle}
\usepackage{amsmath,amsfonts,bbm,mathrsfs}
\usepackage{tabularx} 
\usepackage{epsfig}

\begin{document}
\title{\vskip-1cm Non-Factorizable
Branes on the Torus}
\author{Harun Omer}
\oneaddress{
Jefferson Physical Laboratory\\
Harvard University\\
Cambridge MA 02138\\
USA\\
    {\tt homer@fas.harvard.edu}\\{~}\\}

 \nobreak
\Abstract{
This work discusses string compactifications on the torus with optional $\mathbb{Z}_4 \times \mathbb{Z}_4$ or $\mathbb{Z}_2 \times \mathbb{Z}_2$ orbifold action from the perspective of matrix factorizations.
The method is brought to a level where model building on these backgrounds is possible. Whereas branes discussed in the literature
typically wrap factorizable cycles, that is, cycles in $H^1(\mathbbm{T}^2,\mathbbm{Z})^3 \subset H^3(\mathbbm{T}^6,\mathbbm{Z})$,
branes studied here can be in generic homology classes, can have arbitrary position and Wilson line, have full complex structure respectively K\"ahler moduli dependence and can be subject to any consistent orientifold action. It is shown how any desired D-brane can be constructed systematically.
Three-point correlators can be computed as is demonstrated at hand of an example. Their normalization is not discussed. 
}

\date{November 2008}
\maketitle
\tableofcontents
\pagebreak

\section{Introduction}
This work discusses toroidal orbifolds and orientifolds from the somewhat unconventional perspective of matrix factorizations.
Matrix factorizations are a way to describe topological D-branes as sheaves, which is a significantly more
comprehensive description in comparison to most model building work with intersecting branes
where D-branes are typically reduced to their homology charge only. Notably the complex structure as well as open string moduli are encoded.
Apart from the significance of the reformulation in its own right, the results should
also be interesting from a model building perspective. In the literature of intersecting branes on toroidal type IIA backgrounds the branes
considered are normally only factorizable branes, that is branes wrapping a cycle in $H^1(\mathbbm{T}^2,\mathbbm{Z})^3 \subset H^3(\mathbbm{T}^6,\mathbbm{Z})$.
Due to its simplicity but also its special features and maximal symmetries, the torus has been an
important model case, with countless papers written about it. Nevertheless virtually every
D-brane that appears in these papers is of factorizable type. In fact, in 
the model-building literature D-branes are typically reduced to no more than a charge vector.
As soon as the restriction for the brane to wrap a product of 1-cycles is lifted,
little can be done. Is the brane stable? Do the branes recombine? Does a brane decay into
other branes? Is the configuration still supersymmetric? These and other questions
can not be answered easily for a generic brane on the $\mathbb{T}^6$, not to mention the dynamics
of low-energy physics. See for instance~\cite{nonfact1,nonfact2} for two of the rare papers on tori where a part of these restrictions are lifted.
Another issue is the difficulty of obtaining an odd number of families
in the important $\mathbbm{T}^6/ \mathbbm{Z}_2 \times \mathbbm{Z}_2$-orbifolds.
Non-factorizable branes can give rise to odd numbers of families without resorting
to tilted tori.
Furthermore, factorizable branes can only lead to intersection numbers of
three when the branes intersect three times on one two-torus and once on each of the
other. As a result, the intricate process of selecting branes in a manner that
the standard model gauge group and the chiral spectrum is reproduced 
and the tadpole cancellation conditions are satisfied, leads to models whose
D-brane configuration always seems arbitrary. After all, from the type II perspective,
obtaining precisely a multiplicity of three for {\it all} intersections is
a striking 'coincidence'. It would seem more natural to suspect that the
brane configuration should reflect the symmetry of this coincidence. Perhaps
non-factorizable D-branes could allow a more
symmetric brane configuration than the artificially-seeming factorizable constructions. 
With the method used in this work there is no restriction to factorizable cycles only.
\\
Another important issue, in fact one of the main open problems in string theory,
is the lack of a practically feasible systematic way to obtain Yukawa couplings in generic situations.
Yukawa couplings can currently be computed for branes on a two-dimensional torus by
exponentiating and summing up the instanton areas enclosed by the intersecting branes~\cite{yukawaT2}. Computations on higher-dimensional tori have been only performed for factorizable branes and yet again for generic branes computations cease to be feasible in practice. 
This paper is also intended as a step in the direction of computing Yukawa couplings at least on toroidal backgrounds since generic branes can be described and three-point correlators can be computed. This is demonstrated at hand of an example. In order to compare with experiment, these Yukawa couplings still have to be normalized properly which is not done here.
\\\\
In this paper together with the preceeding paper~\cite{harunT6}, all necessary
model-building ingredients are introduced
for the $\mathbb{T}^2 \times \mathbb{T}^2 \times \mathbb{T}^2$ with the optional choice
of a $\mathbb{Z}_4 \times \mathbb{Z}_4$ or a $\mathbb{Z}_2 \times \mathbb{Z}_2$-
orbifold action. Any consistent orientifold action can be imposed in addition.
It is shown that every BPS-brane on these backgrounds has an analog
in matrix factorizations and it is explained how the factorization of any
given D-brane can be constructed systematically. In principle the quartic-curve
Landau-Ginzburg potential used here could be replaced with
the potential of the $\mathbb{Z}_3$-orbifold or the $\mathbb{Z}_6$ orbifold as well.
Once a model has been set up, the three-point functions can be computed by the Kapustin-Li
formula~\cite{kapustinli}. Higher point functions are not discussed here, see~\cite{knappomer1} for
approaches to compute them.
It should also be noted that in order to derive the CKM-matrix, it would be necessary
to normalize the correlators, for instance by imposing the analog of Picard-Fuchs equations, whose open string version is unfortunately not yet known.
This gap is admittedly a drawback. Nevertheless, the dependence of the correlators on the complex structure as
well as open string moduli can still be seen.  
The purpose of this paper is to demonstrate how model building for the torus works with matrix factorizations and in full generality
and the example is not designed to reproduce our real-world CKM-matrix. It should be understood
as a basis for model-building, demonstrating that it is feasible even for 
models complex enough to have a Pati-Salam model or MSSM spectrum.
Furthermore, some hurdles have to be overcome to apply these techniques to other
backgrounds with Landau-Ginzburg description, notably the problem of chosing the right
parametrization for the moduli, a problem that is solved in the special case of the torus by the 
parametrization given in section 3. Nevertheless it is hoped that this work proves as a helpful basis for other aspects in applying this method
to other backgrounds.
\\\\
The material is organized as follows. Section~\ref{sec:lgmodel} introduces the Landau-Ginzburg
description of the toroidal models discussed and section~\ref{sec:moduli} discusses the moduli dependence.
The paragraphs~\ref{sec:matfac}-\ref{sec:T6Z4xZ4} and \ref{sec:moduli1}-\ref{sec:moduli2}
are review material and serve as introduction and to fix the notation.
These sections overlap with~\cite{harunT6}, the rest of the material presented is new. See also~\cite{harunT6} for
issues of stability of the configuration (and~\cite{Walcher:2004tx} for brane stability). Section~\ref{sec:homclass} discusses how the homology class can be extracted from a given matrix factorization
and how a matrix factorization is obtained for a brane of any given homology class.\\\\
The word 'factorizable' is used in different contexts in the literature and 
a few words are in order here to avoid confusion. The toroidal manifold can be 'factorizable',
if the $T^6$ factors into $T^2 \times T^2 \times T^2$. This is always the case in this discussion.
On this background, the type IIA 3-cycles may or may not be factorizable. They are called 'factorizable' when
they are products of 1-cycles, each wrapping one $T^2$ and this is the meaning of the word in the context of this paper.
Finally, 'matrix factorizations' are yet again something completely different. There, 'factorization' comes from the fact that the Landau-Ginzburg superpotential is factored into matrices.
\section{The LG description of toroidal orbifolds}
\label{sec:lgmodel}
\subsection{Matrix factorizations}
\label{sec:matfac}
A matrix factorization
$Q=\begin{pmatrix} 0 & E\\J & 0 \end{pmatrix}$ is a block off-diagonal matrix with polynomial
entries in the variables of the LG-superpotential $W$ which satisfies the factorization condition,
\begin{eqnarray}
Q^2=W\mathbbm{1} \;\mbox{  or  }\; EJ=JE=W\mathbbm{1}.\label{eq:factcond}
\end{eqnarray}
It encodes the boundary condition of the LG-theory. From a more mathematical perspective it can
be regarded as a realization of the derived category of coherent sheaves. Space constraints 
prohibit a meaningful introductory review of the subject; a partial review about matrix factorizations is~\cite{matfac3}.
Related work is found in~\cite{matfac1,matfac2,orlov1,Orlov:2003yp,Lazaroiu:2003md,Kapustin:2003rc,Ashok:2004zb,Ashok:2004xq,Herbst:2004jp,Herbst:2004ax,Brunner:2004mt,Brunner:2005fv,Enger:2005jk,Dell'Aquila:2005jg,tachcubic,Herbst:2006nn,Walcher:2006rs,Govindarajan:2006uy,Jockers:2006sm,Baumgartl:2007an,Morrison:2007bm,hhp,Krefl:2008sj,Knapp:2008uw,Jockers:2008pe}.
\subsection{The plain $\mathbbm{T}^2$}
The plain $\mathbbm{T}^2$ has been analyzed exhaustively in~\cite{omerknapp2} and the model of the next paragraph,
the $\mathbbm{T}^6/\mathbbm{Z}_4 \times \mathbbm{Z}_4$ orbifold has been discussed in some detail in~\cite{harunT6}.
We need results of the two papers as a basis and the following few lines serve to remind of the model definition and to introduce the notation.\\
Consider a Landau-Ginzburg model with superpotential,
\begin{eqnarray}
W_{T^2}=x_1^4+x_2^4 - a\, x_1^2 x_2^2 - z_1^2,\label{eq:supotT2}
\end{eqnarray}
where the square term $z_1^2$, which can be integrated out in the action of a bulk-only theory, was added in order to achieve the right fermion number. The equation $W=0$ describes a torus in weighted projective space $\mathbbm{P}^4_{1,1,2}$.
The superpotential is invariant under a $\mathbbm{Z}_4$ symmetry,
\begin{eqnarray}
g_0:\;(x_1,x_2,z_1) \mapsto (i x_1,i x_2,-z_1),
\end{eqnarray}
which can be divided out of the theory. In that case the large volume limit of the theory has an interpretation as a torus $\mathbbm{T}^2$ whose complex structure $\tau$ is parametrized by $a=a(\tau)$. The A-side mirror is a rigid $\mathbbm{T}^2$ with K\"ahler structure $\tau$. 
\subsection{The $\mathbbm{T}^6/\mathbbm{Z}_4 \times \mathbbm{Z}_4$ orbifold}
\label{sec:T6Z4xZ4}
In a tensor product of three such theories,
\begin{eqnarray}
W=\sum_{i=1}^6 x_i^4 - a_1\, x_1^2 x_2^2 - a_2\, x_3^2 x_4^2 - a_3\, x_5^2 x_6^2 - z_1^2\label{eq:WT6} \;\;\;\;\;(-z_2^2-z_3^2),
\end{eqnarray}
each of the three building blocks contributes one $\mathbbm{Z}_4$-symmetry, giving rise to a total $(\mathbbm{Z}_4)^3$ symmetry. If we divide out a diagonal phase rotation,
\begin{eqnarray}
g_0:\;(x_1,x_2,x_3,x_4,x_5,x_6,z_1) \mapsto (i x_1,i x_2,i x_3,i x_4,i x_5,i x_6,-z_1),
\end{eqnarray}
as we did for the $\mathbbm{T}^2$, the large-volume interpretation of the resulting model has an interpretation as a $\mathbbm{T}^6/\mathbbm{Z}_4 \times \mathbbm{Z}_4$-orbifold.
An arbitrary even number of additional square terms can be added to the superpotential with no effect on the physical theory. 
Adding two squared terms to $W$ as indicated by the parentheses in Eq.~(\ref{eq:WT6}) has the advantage that the connection to a tensor theory of three $W_{\mathbbm{T}^2}$ becomes more evident. Matrix factorizations describing branes on the $\mathbbm{T}^2$ -- their cycles are in the homology class $H^1(\mathbbm{T}^2,\mathbbm{Z})$ on the A-side -- can then be tensored to describe a product brane with homology class $H^1(\mathbbm{T}^2,\mathbbm{Z})^3 \subset H^3(\mathbbm{T}^6,\mathbbm{Z})$ in the tensored theory.\\\\
Note that the orbifolding is implicit in the theory by construction, which means that one single factorization describes the brane together with all of its orbifold images. Therefore, the direct product of three branes, each wrapping one of the tori, is a factorizable brane of these three cycles plus its orbifold images. This is different from a possible orientifolding: An image brane of the orientifold action is described by an additional matrix factorization. This will be discussed further below. 
\subsection{The rigid $\mathbbm{T}^6$}
When the entire symmetry $(\mathbbm{Z}_4)^3$ is treated as a quantum symmetry, all three group generators,
\begin{eqnarray}
\begin{array}{l}
g_1:\;(x_1,x_2,x_3,x_4,x_5,x_6,z_1,z_2,z_3) \mapsto (i x_1,i x_2,x_3,x_4,x_5,x_6,-z_1,z_2,z_3)\\
g_2:\;(x_1,x_2,x_3,x_4,x_5,x_6,z_1,z_2,z_3) \mapsto (x_1,x_2,i x_3,i x_4,x_5,x_6,z_1,-z_2,z_3)\\
g_3:\;(x_1,x_2,x_3,x_4,x_5,x_6,z_1,z_2,z_3) \mapsto (x_1,x_2,x_3,x_4,i x_5,i x_6,z_1,z_2,-z_3),
\end{array}
\end{eqnarray}
are divided out instead of just one diagonal phase rotation. This corresponds to a product of three $\mathbbm{T}^2$. In such a theory, every factorization carries three labels $n_{1,2,3}=0,1$, which are associated with the three generators $g_{1,2,3}$ respectively and can take two different values. Altogether there are therefore $2^3=8$ group-elements which can be arranged in the order $\tilde g_{n_1+2n_2+4n_3+1}=g_1^{n_1} g_2^{n_2} g_3^{n_3}$, or, more explicitly:
\begin{eqnarray}
(\tilde g_1,\tilde g_2,\tilde g_3,\tilde g_4,\tilde g_5,\tilde g_6,\tilde g_7,\tilde g_8):=(1,g_1,g_2,g_1 g_2,g_3,g_1 g_3,g_2 g_3,g_1 g_2 g_3).\label{eq:groupordering}
\end{eqnarray}
On the IIA side, the effect of the group generators $g_{j}$ can be interpreted as a rotation by $\pi/2$ on the $j$-th $\mathbbm{T}^2$ in analogy to what is known from the $\mathbbm{T}^2$~\cite{omerknapp2}.
\subsection{The $\mathbbm{T}^6/\mathbbm{Z}_2 \times \mathbbm{Z}_2$ orbifold}
Since $\mathbbm{Z}_2 \times \mathbbm{Z}_2$ is a subgroup of $\mathbbm{Z}_4 \times \mathbbm{Z}_4$, the $\mathbbm{T}^6/\mathbbm{Z}_2 \times \mathbbm{Z}_2$-orbifold can be obtained by retaining two $\mathbb{Z}_2$ subgroups in the quantum group of the LG theory.
To begin with, one $\mathbbm{Z}_4$-quantum symmetry generated by,
\begin{eqnarray}
g_0:\;(x_1,x_2,x_3,x_4,x_5,x_6,z_1) \mapsto (i x_1,i x_2,i x_3,i x_4,i x_5,i x_6,-z_1),
\end{eqnarray}
is modded out of the full $(\mathbbm{Z}_4)^3$-symmetry. On the IIA side the action of $g_0=g_1 g_2 g_3$ can be interpreted as a rotation by $\pi/2$ on each $\mathbbm{T}^2$. To obtain the desired subgroup $\mathbbm{Z}_2 \times \mathbbm{Z}_2$, we define the additional quantum orbit generators $h_1=g_1^2 g_2^{-2}$ and $h_2=g_2^2 g_3^{-2}$:
\begin{eqnarray}
\begin{array}{l}
h_1:\;(x_1,x_2,x_3,x_4,x_5,x_6,z_1) \mapsto (-x_1,-x_2,-x_3,-x_4,x_5,x_6,z_1),\\
h_2:\;(x_1,x_2,x_3,x_4,x_5,x_6,z_1) \mapsto (x_1,x_2,-x_3,-x_4,-x_5,-x_6,z_1).
\end{array}
\end{eqnarray}
A theory invariant under these generators has an orbifold group that is reduced further down from the remaining $(\mathbbm{Z}_4)^2$-symmetry to the desired $\mathbbm{Z}_2 \times \mathbbm{Z}_2$ subgroup.\\\\
Turning on further deformations in the LG potential corresponds to blowing-up the orbifold singularities. Such blow-ups are not considered here since the focus of the paper is the computation of correlation functions.
\subsection{Orbifolds}
The orbifold condition on matrix factorizations is well known for the orbifold group associated with the $U(1)$ R-symmetry of the theory~\cite{Walcher:2004tx}, which is generated by $g_0$. But here we have to deal with other group generators such as $g_{1,2,3}$ and $h_{1,2}$. What is the phase of a factorization in such an orbit? What is its large volume interpretation? In order to answer these questions, the orbifold condition is revisited here.\\
The Calabi-Yau manifolds which have a description by a Landau-Ginzburg theory are the vanishing locus of a polynomial $W$ of total degree $N$ in weighted projective space $\mathbbm{P}^{r-1}_{w_1,...,w_r}$ where $\sum_{i=1}^r w_i=N$. The CY/LG-correspondence associates this CY to a LG theory with superpotential $W$, orbifold group $\mathbbm{Z}_N$ and central charge $\hat c=r-2$. The variables in the LG potential have $R$-charges $q_i=2 w_i/N$.
The R-symmetry transforms the measure of the integral which is cancelled by the quasi-homogeneity condition,
\begin{eqnarray}
W(e^{i\lambda q_i} x_i)=e^{2i\lambda}W(x_i)\qquad \forall \lambda \in \mathbbm{R},
\end{eqnarray}
of the superpotential. The R-symmetry further descends to the factorization by a generator $\rho(\lambda)$ that satisfies,
\begin{eqnarray}
\rho(\lambda)Q(e^{i\lambda q_i} x_i)\rho^{-1}(\lambda)=e^{i\lambda}Q(x_i).\label{eq:rho}
\end{eqnarray}
The corresponding $U(1)$ vector field is given by,
\begin{eqnarray}
R(\lambda)=-i \partial_{\lambda} \rho(\lambda,x_i)   \rho^{-1}(\lambda,x_i),
\end{eqnarray}
which turns out to be actually independent of $\lambda$. Eq.~(\ref{eq:rho}) can now be rewritten as,
\begin{eqnarray}
EQ+[R,Q]=Q\mbox{  with  } E=\sum_i q_i x_i\frac{\partial}{\partial x_i}.\label{eq:chQis1}
\end{eqnarray}
From the R-matrix one obtaines the orbifold generator $\gamma$,
\begin{eqnarray}
\gamma(g)=\mbox{diag}(1,....,1,-1,...,-1)e^{i \pi R} e^{-\pi i \phi(g)}.\label{eq:orb}
\end{eqnarray} 
The orbifold condition for a group element $g$ on a matrix factorization $Q(x)$ reads,
\begin{eqnarray}
\gamma(g)\,Q(g x)\, \gamma(g)^{-1} = Q(x),\label{eq:orbi}
\end{eqnarray}
where the orbifold matrix $\gamma(g)$ is subject to the constraint $\gamma(g)^N=1$. It is this constraint that singles out $N$ angles $\phi(g)$ in Eq.~(\ref{eq:orb}), which are the phases of the brane in the corresponding orbit $g$. 
The analog of an R-matrix, however, can not be set-up for arbitrary group actions since the derivation relies heavily on the quasi-homogeneity. Using Eq.~(\ref{eq:orbi}) one can find the group generators $\gamma(g)$, but it is not clear what phase the brane has, since the equation is invariant under $\gamma(g) \mapsto e^{i \alpha} \gamma(g)$ for arbitrary phases $\alpha$.
We need to impose an additional condition in order to fix that phase. In the particular case of the $T^6$ one ansatz works as follows. First, we obtain an $R_g(\alpha)$ for a given orbifold matrix from Eq.~(\ref{eq:orb}). We group the variables in the LG potential into one class on which $g$ acts trivially, $x_1,...,x_M$ and one class on which it acts non-trivially, $x_{M+1},...,x_N$. On $R_g(\alpha)$ we impose Eq.~(\ref{eq:chQis1}), restricted to the non-trivial variables only:
\begin{eqnarray}
E Q_g+R_g Q_g - Q_g R_g=Q_g\mbox{   where   }Q_g=Q|_{x_1=0,...,x_M=0}.
\end{eqnarray}
In other words, we use the normalization of the diagonal phase and project the generators onto it. But it is possible to avoid the R-matrix
altogether and set up the orbifold generators directly. They are determined up to the phase ambiguity $\gamma(g) \mapsto e^{i \alpha} \gamma(g)$.
The orbifold condition on an open string $\Psi(x)$ between two factorizations $Q$ and $Q'$ with orbifold generators $\gamma(g)$, $\gamma'(g)$ is,
\begin{eqnarray}
\gamma(g)\,\Psi(g x)\, \gamma'(g)^{-1} = e^{i \Delta\alpha}\Psi(x).\label{eq:orbimor}
\end{eqnarray}
The angle $\Delta \alpha$ is the difference of the phases of supersymmetry of the branes and given the phase of one brane the phase of the second
brane is  uniquely determined.
The total phase of a brane in an orbit $g=g_0^{l_1} h_1^{l_2} h_2^{l_3}$ is simply the sum of the phases associated with each orbit label $l_{1,2,3}$.
Here, a change in the $g_0$-orbit changes the phase by $\pi/2$ whereas the $h_1$ or $h_2$-orbits differ by a phase of $\pi$ each.
In fact, this is already enough information to understand the geometric meaning of the orbits of $h_{1,2}$ on the $A$-side.
A phase difference of $\pi/2$ corresponds to a rotation by $\pi/2$ on one torus~\cite{omerknapp2} in the $A$-side picture.
The group action of $h_1$ treats two tori on equal footing and singles out the third, therefore different $h_1$-orbits will either differ by a rotation on both the first and the second torus, or only by a rotation on the third torus.
The phase $\pi$ which is picked up by a change of orbit in $h_{1,2}$ should be a rotation on two tori so it is clear that the action of $h_1$ rotates the brane by $\pi/2$ on the first and second torus in the $A$-side picture. By the same argument, $h_2$ rotates a cycle on the second and third torus.\\
The branes in the $\mathbbm{T}^6 /\mathbb{Z}_2 \times \mathbb{Z}_2$ are of course nothing but branes on the plain $\mathbbm{T}^6$ plus their orbifold images (with the same homology charge) and from what has been argued, the connection in terms of group elements can be established as follows:
\begin{eqnarray}
(1,g_0 h_2,g_0 h_1 h_2,h_1,g_0 h_1,h_1 h_2,h_2,g_0)\simeq (1,g_1,g_2,g_1 g_2,g_3,g_1 g_3,g_2 g_3,g_1 g_2 g_3).\label{eq:Z2xZ2toT6}
\end{eqnarray}
Remember that $g_j$ generates a rotation by $\pi/2$ on the $j$-th torus on the A-side. Using that information and the identification Eq.~(\ref{eq:Z2xZ2toT6}) we can visualize the A-side D-branes. Further below we will also use will also use Eq.~(\ref{eq:Z2xZ2toT6}) to apply the same transformation of the plain $\mathbbm{T}^6$ to its orbifolded version when converting the Landau-Ginzburg $R$-charges to the large-volume charges.\\\\
The $R$-charges can be computed from the orbifold matrices $\gamma$ in a straightforward manner in a one-generator model~\cite{Walcher:2004tx}.
As mentioned before, there is also a phase associated with each brane. It is inherited from the grading of the category. The eigenvalue equation,
\begin{eqnarray}
E \Psi+R' \Psi-\Psi R=q \Psi
\end{eqnarray}
assigns a charge $q$ to each morphism in $\mbox{Hom}(Q,Q')$. The charge equals the difference of the phases of the branes between which it maps.
For a one-generator theory, the general formula for the ovelap between a some $n$-th group representative of a brane $Q$ with the $k$-th Ramond ground state in the twisted sector is, 
\begin{eqnarray}
\langle Q^{(n)} | k\rangle =\mbox{STr}[\, \gamma(g)^k] i^{k\cdot\, n}.\label{eq:T2charge}
\end{eqnarray}
The two choices for $n$ and two for $k$ can be grouped together in a $2\times 2$-matrix.
For a brane on the $\mathbbm{T}^2 \times \mathbbm{T}^2 \times \mathbbm{T}^2$ we generalize the charge formula to be,
\begin{eqnarray}
\langle Q^{(n_1,n_2,n_3)} | k_1,k_2,k_3 \rangle =\mbox{STr}[\, \gamma(g_1)^{k_1} \gamma(g_2)^{k_2} \gamma(g_3)^{k_3}]\;i^{\,k_1\,n_1}\;i^{\,k_2  \,n_2}\;i^{\,k_3\,n_3}.  \label{eq:Rchargeformula}
\end{eqnarray}
Sorting the $2^3=8$ choices for the branes in the orbits labeled by $n_i$ and the same number of possibilities for the twisted sectors $k_i$ according to the conventions of Eq.~(\ref{eq:groupordering}), we can set up a $\mathbb{C}^{8\times 8}$ R-charge matrix for every factorization. For a $\mathbbm{T}^2 \times \mathbbm{T}^2$, we need only the $2^2=4$ generators $\tilde g_1,...,\tilde g_4$ so the charges $\langle Q^{(n_1,n_2)} | k_1,k_2 \rangle$ can be stored in a $\mathbb{C}^{4\times 4}$-matrix. The formula Eq.~(\ref{eq:Rchargeformula}) is obviously correct for a tensor product of branes on $\mathbbm{T}^2$ since it is simply a product composed of factors of Eq.~(\ref{eq:T2charge}), but the formula is by no means restricted to branes which can be decomposed as a product of branes on $\mathbbm{T}^2$. Rather the purpose of the construction is precisely to deal with the class of non-factorizable branes.
\section{Moduli dependence}
\label{sec:moduli}
A complete description of branes on toroidal backgrounds should take into account the complex structure and K\"ahler moduli as well as all open string moduli.
This section contains nothing new. It just recaps some results of~\cite{harunT6} without proof.
From the IIB perspective, there are the three complex structure moduli of the three $\mathbbm{T}^2$, which are the K\"ahler parameters of the square tori on the IIA side. In addition, there are open string moduli. A torus $\mathbbm{T}^2$ with complex structure $\tau$ has an open string moduli space $\mathcal{M}=\mathbb{C}/(\mathbb{Z}+\tau \mathbb{Z})$. From the IIB-perspective, a point $u\in \mathcal{M}$ on the moduli space defines the location of the D0-brane component on the torus. On the IIA-side, the brane is mapped into a D1-brane so that instead of a complex number only a real number is needed to mark its location on the torus. The boundary modulus can be decomposed into $u\equiv u^{\parallel}+\tau u^{\perp}$ and contains the location of the $D1$ brane in the real component $u^{\parallel}$ in terms of the distance of the brane to the origin. The second real number $u^{\perp}$ gives the value of the Wilson line of the brane.\\
D-brane categories or their realization as matrix factorizations are able to encode all these moduli: The complex structure, the location and the Wilson line. In order to extract this information, it is necessary to use flat coordinates, which for the torus amounts to parametrizing the moduli in terms of the Jacobi theta-functions $\Theta_{1,...,4}$ as was done in~\cite{omerknapp2}. The complex structure modulus $a=a(\tau)$ reads,
\begin{eqnarray}
a(\tau)=\frac{\Theta_2^4(2\tau)+\Theta_3^4(2\tau)}{\Theta_2^2(2\tau)\Theta_3^2(2\tau)}.
\end{eqnarray}
It satisfies,
\begin{eqnarray}
W_{T^2}(x_1=\alpha_1,x_2=\alpha_2,z_1=\alpha_3)=\alpha_1^4+\alpha_2^4-a\; \alpha_1^2 \alpha_2^2 - \alpha_3^2=0, \label{eq:thetarel}
\end{eqnarray}
where
\begin{eqnarray}
\begin{array}{l}
\displaystyle \alpha_1(u,\tau)=\Theta_2(2u,2\tau)\qquad \alpha_2(u,\tau)=\Theta_3(2u,2\tau),\\
\displaystyle \alpha_3(u,\tau)=\frac{\Theta_4^2(2\tau)}{\Theta_2(2\tau)\Theta_3(2\tau)}\Theta_1(2u,2\tau)\Theta_4(2u,2\tau).
\end{array}
\end{eqnarray}
Solving Eq.~(\ref{eq:thetarel}) for $a(\tau)$ in terms of $\alpha_i$, we can work with this expression for $a(\tau)$ when factorizing the Landau-Ginzburg potential into matrices. The purpose is that for any given complex structure $\tau$ we now have an entire family of factorizations parametrized by the argument $u$ in the functions $\alpha_i(u,\tau)$. These functions are actually sections of line bundles.\\
\subsection{Branes on the $\mathbbm{T}^2$}
\label{sec:moduli1}
Two factorizations $Q_{a,b}$ of the $\mathbbm{T}^2$ superpotential are,
\begin{eqnarray}
\begin{array}{l}
E_a=\begin{pmatrix}
X_1 & d_1 x_1 x_2 + z_1\\
d_1 x_1 x_2 - z_1 &  -X_2 X_3 X_4
\end{pmatrix}\qquad
J_a=\begin{pmatrix}
X_2 X_3 X_4 & d_1 x_1 x_2 + z_1\\
d_1 x_1 x_2 - z_1 & -X_1
\end{pmatrix}\\
\label{eq:T2branes}\\
E_b=\begin{pmatrix}
X_1 X_2 & d_1 x_1 x_2 + z_1\\
d_1 x_1 x_2 - z_1 &  -X_3 X_4
\end{pmatrix}\qquad
J_b=\begin{pmatrix}
X_3 X_4 & d_1 x_1 x_2 + z_1\\
d_1 x_1 x_2 - z_1 & -X_1 X_2
\end{pmatrix}
\end{array}
\end{eqnarray}
Here a simplifying notation was introduced:
\begin{eqnarray}
\displaystyle X_1=(x_1+c_1 x_2)\qquad X_2=(x_1-c_1 x_2)\qquad X_3=(x_1+\frac{1}{c_1} x_2)\qquad X_4=(x_1-\frac{1}{c_1} x_2),
\end{eqnarray}
with,
\begin{eqnarray} 
\begin{array}{l}
\displaystyle c_i=\frac{\alpha_2(u_i,\tau_i)}{\alpha_1(u_i,\tau_i)}\qquad
\displaystyle d_i=\frac{\alpha_3(u_i,\tau_i)}{\alpha_1(u_i,\tau_i) \alpha_2(u_i,\tau_i)}
\end{array}\label{eq:ci}
\end{eqnarray}
Explicit multiplication shows that both satisfy the factorization condition Eq.~(\ref{eq:factcond})
where Eq.~(\ref{eq:thetarel}) ensures that this is true for any value of the background geometry modulus $\tau_i$ of the torus and the boundary modulus $u_i$. In a tensor product we replace $X_k$ with $Y_k$ and $Z_k$ on the second and third torus respectively:
\begin{eqnarray}
\displaystyle Y_1=(x_3+c_2 x_4)\qquad Y_2=(x_3-c_2 x_4)\qquad Y_3=(x_3+\frac{1}{c_2} x_4)\qquad Y_4=(x_3-\frac{1}{c_2} x_4),\\
\displaystyle Z_1=(x_5+c_3 x_6)\qquad Z_2=(x_5-c_3 x_6)\qquad Z_3=(x_5+\frac{1}{c_3} x_6)\qquad Z_4=(x_5-\frac{1}{c_3} x_6).
\end{eqnarray}
Likewise $z_1$ becomes $z_2$ or $z_3$. When we are talking about the second torus, $E_a$ and $J_a$ for example would stand for:
\begin{eqnarray}
\begin{array}{l}
E_a=\begin{pmatrix}
Y_1 & d_2 x_3 x_4 + z_2\\
d_2 x_3 x_4 - z_2 &  -Y_2 Y_3 Y_4
\end{pmatrix}\qquad
J_a=\begin{pmatrix}
Y_2 Y_3 Y_4 & d_2 x_3 x_4 + z_2\\
d_2 x_3 x_4 - z_2 & -Y_1
\end{pmatrix}
\end{array}\label{eq:T2branes2}
\end{eqnarray}
To avoid overloading the notation, the factorizations carry no label numbering the tori and the explicit dependence on fields and moduli is suppressed. The fields and parameters are associated with the three tori as follows:\\\\
\begin{tabular}{c|c}
Torus & fields and moduli\\
\hline
$\mathbb{T}^2_1$: & $x_1$,$x_2$\qquad $c_1$,$d_1$\qquad $z_1$\qquad $X_k$\\
$\mathbb{T}^2_2$: & $x_3$,$x_4$\qquad $c_2$,$d_2$\qquad $z_2$\qquad $Y_k$\\
$\mathbb{T}^2_3$: & $x_5$,$x_6$\qquad $c_3$,$d_3$\qquad $z_3$\qquad $Z_k$
\end{tabular}
\\\\The notation $Q_a \otimes Q_a$ for example denotes the product of the factorization defined by Eq.~(\ref{eq:T2branes}) with the one in Eq.~(\ref{eq:T2branes2}).
\subsection{Branes on the $\mathbbm{T}^6/\mathbbm{Z}_4 \times \mathbbm{Z}_4$ orbifold}
In~\cite{harunT6} three branes $Q_a$ were tensored together. There is only a single orbifold label from the quantum $\mathbbm{Z}_4$-symmetry. It was shown that the two orbits correspond to the two factorizable cycles,
\begin{eqnarray}
\Pi_1 = 4(\pi_{135} - \pi_{236} -\pi_{146}-\pi_{245}),\\
\Pi_2 = 4(\pi_{136} - \pi_{235} -\pi_{145}-\pi_{246}),
\end{eqnarray}
where the standard model-building notation is used which denotes the two
fundamental cycles on the n-th torus as $\pi_{2n-1}$ and $\pi_{2n}$ and uses the definition,
\begin{eqnarray}
\pi_{klm}\equiv \pi_k \otimes \pi_l \otimes \pi_m.
\end{eqnarray}
The cycles $\Pi_1$ and $\Pi_2$ differ by a rotation of $\pi/2$ on each of the three $\mathbbm{T}^2$.
Again, the moduli $u^{\parallel}_i$ denotes the distance the brane is shifted from the origin of the $i$-th torus, $u^{\perp}_i$ is the value of the Wilson line along it and $\tau_i$ is its the K\"ahler structure on the A-side. On the B-side, $\tau_i$ are of course the complex structures of the tori and the tensored factorizations correspond to branes with fluxes where in the topological model the branes with flux can as usual be interpreted as bound states of the branes with lower dimensional brane components whose location is parametrized by the $u_i$.
\subsection{Branes on the $\mathbbm{T}^6$}
\label{sec:moduli2}
When dealing with a plain $\mathbbm{T}^6$, we have three quantum numbers $n_1,n_2,n_3$ taking the values $0$ or $1$. In the IIA picture, each label is associated with a rotation by $\pi/2$ around one $\mathbbm{T}^2$. The brane $(Q_a \otimes Q_a \otimes Q_a)^{(n_1,n_2,n_3)}$ therefore denotes a brane with homology class $\pi_{1+n_1,3+n_2,5+n_3}$. The discussion concerning the moduli does not change.
\subsection{Non-factorizable cycles}
The branes discussed above wrap cycles which are factorizable on the IIA side, which means that they are a direct product of three 1-cycles. Again it must be emphasized that this notion of "factorizable" branes is standard in the type II model building literature and has nothing to do with the "factorization" in "matrix factorization".\\
The real strength of matrix factorizations is that they are capable of encoding general boundary conditions. Non-factorizable cycles pose serious difficulties to conventional intersecting brane modeling methods. It is not unfair to say that even in the case of the torus hardly any work at all has been done on them. Brane recombination processes are generally reduced to the addition of K-theory charges. This is good enough to determine the spectrum of a model and verify tadpole cancellation but for verifying supersymmetry or computing Yukawa couplings or the effective superpotential
more refined ways to describe branes are needed.\\
Matrix factorizations can be set-up directly by an educated guess or can be obtained by tachyon condensation from other branes, followed by a change in
the open string moduli. In the latter case, one starts with a minimal set of branes which have a unimodular intersection number. By bound state formation one can generate D-branes in any desired homology class of the torus orbifolds. This can be realized as follows. The minimal set of branes is here given by the eight fundamental cycles which are the eight orbits of the factorization $(Q_a \otimes Q_a \otimes Q_a)^{(n_1,n_2,n_3)}$. By writing the homology charge as a linear combination of the homology charge of the branes in the minimal basis, the basis branes which have to be recombined to obtain the desired brane can be established. A charge vector $\vec{q}=(1,0,2,0,0,0,0,0)$ for example can be decomposed as,
\begin{eqnarray}
\begin{array}{rcl}
\vec{q}&=&(1,0,2,0,0,0,0,0)\\
&=&(1,0,0,0,0,0,0,0)+(0,0,1,0,0,0,0,0)+(0,0,1,0,0,0,0,0),
\end{array}
\end{eqnarray}
where factorizations carrying the charges on the right hand side are just the basis branes mentioned above and discussed in~\cite{harunT6} as well as in the next section.
Using tachyon condensation (see e.g.~\cite{tachcubic}) one representative in the desired homology class can be constructed. The open string moduli of the resulting brane
depends on the moduli of the branes out of which it has been obtained but it can now be changed easily. The minimal basis is made up of fractional branes and the condensed brane is again a fractional brane passing through fixed points. 
By permuting $\{X_1,X_2,X_3,X_4\}$, $\{Y_1,Y_2,Y_3,Y_4\}$ or $\{Z_1,Z_2,Z_3,Z_4\}$ among each other one permutes the fixed points through which the brane passes and turns Wilson lines on and off. Any factorization obtained by tachyon condensation from the basis branes passing through a maximum number of fixed points can be written as a direct product of some factorization with $(-z_1)(z_1)$. The entire moduli space is a product of three tori. An arbitrary value on one of these tori, say the first one, is achieved by the continous deformation $(-z_1)(z_1) \mapsto (d_1 x_1 x_2-z_1)(d_1 x_1 x_2+z_1)$. A generic bulk brane is obtained by taking a direct sum of two such fractional branes and then continously moving it into the bulk. The moving into the bulk corresponds to turning on finite values for $d_2$ and $d_3$ in the direct sum:
\begin{eqnarray}
\begin{array}{l}
E_{bulk}=\begin{pmatrix}
E_{frac} & (d_2 x_3 x_4 + i d_3 x_5 x_6) \mathbbm{1}\\
(d_2 x_3 x_4 - i d_3 x_5 x_6) \mathbbm{1} & J_{frac}
\end{pmatrix},\\\\
J_{bulk}=\begin{pmatrix}
J_{frac} & -(d_2 x_3 x_4 + i d_3 x_5 x_6) \mathbbm{1}\\
(-d_2 x_3 x_4 + i d_3 x_5 x_6) \mathbbm{1} & E_{frac}
\end{pmatrix}.
\end{array}
\end{eqnarray}
The moduli $\tau_i$ and $u_i$ can now both be varied over the entire moduli space without spoiling the factorization condition.
Every brane we know to exist in the large-volume limit therefore has a description in terms of matrix factorizations.
At the Gepner point, matrix factorizations become Recknagel-Schomerus~\cite{RSbranes} or permutation branes~\cite{permbranes} from the CFT perspective. It is interesting to see that this set of branes is sufficient to generate all BPS branes which was not clear from the outset.
\section{Vector Bundles and Wrapping Numbers}
\label{sec:homclass}
In order to identify factorizations with branes in the classical limit, one extracts the rank and chern characters of
the bundles associated with the matrices. Concretely, the R-charges which we have in the form of Eq.~(\ref{eq:Rchargeformula}) need to be converted into their large volume equivalent. For one particular set of branes the correspondence is known. The Calabi-Yau--Landau-Ginzburg correspondence relates this set of branes on the CY to certain LG states. This correspondence suffices to define the appropriate mapping in full generality which can then be applied to arbitrary branes to establish their classical interpretation.\\
First of all, note that for the R-charges alone the deformation of the superpotential is irrelevant and
we can go to the Gepner-point which is at $a(\tau)=0$ in Eq.~(\ref{eq:supotT2}). There, the LG-model is reduced to a tensor product of two level $2$ minimal models (plus a squared term).
The tensor product factorization of these minimal models is,
\begin{eqnarray}
Q_{tensor}=
\begin{pmatrix} 0 & x_1\\x_1^3 & 0 \end{pmatrix}
\otimes
\begin{pmatrix} 0 & x_2\\x_2^3 & 0 \end{pmatrix}
\otimes
\begin{pmatrix} 0 & z_1\\-z_1 & 0 \end{pmatrix},
\end{eqnarray}
and their orbifold generators are,
\begin{eqnarray}
\gamma=i^n \begin{pmatrix} 1 & 0\\0 & i \end{pmatrix}
\otimes
\begin{pmatrix} 1 & 0\\0 & i \end{pmatrix}
\otimes
\begin{pmatrix} 1 & 0\\0 & -1 \end{pmatrix} \qquad n=0,1.
\end{eqnarray}
The R-charge for the $k$-th twisted sector of the $n$-th brane is obtained by taking the supertrace of the generator $\mbox{STr}\;\gamma^{k}$. Note that instead of running over all four orbits $0,...,3$, $n$ takes only two values since the orbits are indentified pairwise in this model, where the $n+2$-th brane is the anti-brane of brane $n$. The charge matrix becomes,
\begin{eqnarray}
(\langle Q_{tensor}^{(n)} | k \rangle)_{nk} = \begin{pmatrix} 4i &-4\\-4i & -4 \end{pmatrix}.\label{eq:Rch}
\end{eqnarray}
The CY/LG correspondence~\cite{orlov1,hhp} relates these branes in the large volume limit to the restriction of $\wedge^k \Omega(n)$ from the ambient projective space to the hypersurface $W=0$. The rank and chern character of the brane in the $n$-th orbit can be computed and compiled in a matrix as well,
\begin{eqnarray}
(\mbox{ch}_k(Q_{tensor}^{(n)}))_{nk}=\begin{pmatrix} 1 &1\\0 & -1 \end{pmatrix}.\label{eq:chern}
\end{eqnarray}
The first row contains the rank $r$ of the bundle which is the D2-brane charge. The second row contains the first chern class $c_1$ or the D0-brane charge. It must however be multiplied by $2$ since the the hyperplane class intersects the quartic curve $W=0$ twice in the projective space. Concretely, $Q_{tensor}^{(n=0)}$ is a pure D2 brane and $Q_{tensor}^{(n=1)}$ is a bound state of a D2-brane and two anti D0-branes.\\
Since the CY/LG correspondence tells us that in the large volume limit Eq.~(\ref{eq:Rch}) corresponds to the bundle data in Eq.~(\ref{eq:chern}), we define a homomorphism $T^{LV}$ relating the two matrices:
\begin{eqnarray}
T^{LV}\begin{pmatrix} 4i &-4\\-4i & -4 \end{pmatrix}=\mbox{diag}(1,2)\begin{pmatrix} 1 &1\\0 & -1 \end{pmatrix}.
\end{eqnarray}
The diagonal matrix $\mbox{diag}(1,2)$ accounts for the factor of two argued for above. The transformation obtained from the defining equation reads,
\begin{eqnarray}
T^{LV}=\frac{1}{8}\begin{pmatrix} -1-i &-1+i\\2 & 2 \end{pmatrix}.
\end{eqnarray}
We can now act with $T^{LV}$ on the charge matrices for the branes $Q_{a,b}$ of Eq.~(\ref{eq:T2branes}). One gets, 
\begin{eqnarray}
\begin{array}{l}
T^{LV}(\mbox{ch}_k(Q_a^{(n)}))_{nk}=\begin{pmatrix} 0 &1\\1 & -1 \end{pmatrix},\qquad
T^{LV}(\mbox{ch}_k(Q_b^{(n)}))_{nk}=\begin{pmatrix} -1 &1\\2 & 0 \end{pmatrix}.
\end{array}
\end{eqnarray}
Again each column is associated with an orbit $n$ of the orbifold symmetry and allows to read off the IIB side wrapping numbers. The factorization $Q_a^{(n=0)}$ is a pure D0 brane, and $Q_a^{(n=1)}$ is a bound state of a D2 brane and an anti-D0 brane. The factorization $Q_b^{(n=0)}$ is a bound state of an anti-D2 brane, and two D0 branes.\\
Matrix factorizations a priori do not distinguish between bundles differing by monodromies. Were one to trace a loop from the large radius limit around the Gepner point, the chern number of a rank $r$ bundle would change by $\pm 2r$ but the brane would be described by the same factorization.
It must be remembered that the above charges are the large radius charges which makes sense from a B-side perspective since this is a semi-classical situation.\\\\
In order to interpret $(r,c_1)$ as wrapping numbers on the A-side, we need the charges at the Gepner point. By going to the Gepner point, the charges flow according to,
\begin{eqnarray}
(r,c_1) \rightarrow (r,c_1+r).\label{eq:Aside}
\end{eqnarray}
This is equivalent to tensoring by a rank $r$ bundle and is essentially 'half' the monodromy. We therefore need to define a further homomorphism $T^A$ implementing Eq.~(\ref{eq:Aside}) to obtain the IIA-side interpretation:
\begin{eqnarray}
T^A=\begin{pmatrix} 1 &0\\1 & 1 \end{pmatrix}
\end{eqnarray}
For the above branes we get:
\begin{eqnarray}
\begin{array}{l}
T^A T^{LV}(\mbox{ch}_k(Q_a^{(n)}))_{nk}=\begin{pmatrix} 0 &1\\1 & 0 \end{pmatrix},\qquad
T^A T^{LV}(\mbox{ch}_k(Q_b^{(n)}))_{nk}=\begin{pmatrix} -1 &1\\1 & 1 \end{pmatrix}.
\end{array}
\end{eqnarray}
The columns contain the A-side wrapping numbers, so the branes wrap the following cycles:
\begin{eqnarray}
\begin{array}{l}
Q_a^{(n=0)}:\; (0,1)\\
Q_a^{(n=1)}:\; (1,0)\\
Q_b^{(n=0)}:\; (-1,1)\\
Q_b^{(n=1)}:\; (1,1).
\end{array}
\end{eqnarray}
Incrementing the orbifold label $n$ by $1$ corresponds to a rotation by $\pi/2$; a result that was already used above.
The modulus $u^{\parallel}$ in the factorization denotes the distance the brane is shifted from the origin, $u^{\perp}$ is the value of the Wilson line and $\tau$ is the K\"ahler structure on the A-side.\\\\
The transformation matrices $T^A$ and $T^{LV}$ are listed in the appendix for the $\mathbbm{T}^4$ and the $\mathbbm{T}^6$. For any brane $Q$ in some orbit $n_1,n_2,n_3$ one can now compute a charge vector from Eq.~(\ref{eq:Rchargeformula}) and then act with the transformation matrices on it to obtain a vector $\vec{q}$ with the wrapping numbers. Explicitly, in our ordering convention of Eq.~(\ref{eq:groupordering}) the charges $\langle Q|k_1,k_2,k_3\rangle$ enter the charge vector in the following sequence:
\begin{eqnarray}
\begin{array}{l}
\vec{q}_{LG}=
\begin{pmatrix}
\langle Q|1,1,1\rangle\\
\langle Q|3,1,1\rangle\\
\langle Q|1,3,1\rangle\\
\langle Q|3,3,1\rangle\\
\langle Q|1,1,3\rangle\\
\langle Q|3,1,3\rangle\\
\langle Q|1,3,3\rangle\\
\langle Q|3,3,3\rangle
\end{pmatrix}
\end{array}\label{eq:homT6a}
\end{eqnarray}
The homology charges are then given by,
\begin{eqnarray}
\vec{q}=T^A T^{LV} \vec{q}_{LG}.\label{eq:homT6b}
\end{eqnarray}
Of course, $b_3=15$ basis elements would be needed to span $H^3(\mathbbm{T}^6,\mathbbm{Z})$ but the set of BPS branes is smaller. Therefore only eight elements are needed, all others are projected out in the toroidal orbifolds anyway.
The vector notation is convenient to work with and is at times called $\vec{q}$-basis formalism in the intersecting brane literature. The ordering is consistent with the conventions of~\cite{rabadan} which are reproduced in Table~\ref{tab:cyclebasis}.
\begin{table}[t]
\begin{center}
\begin{tabular}{|ccc|}
\hline
basis vector & 3-cycles & wrapping numbers\\
\hline
$q_1$ & $[a] \times [a] \times [a]$ & $n_1 n_2 n_3$ \\
$q_2$ & $[a] \times [a] \times [b]$ & $n_1 n_2 m_3$  \\
$q_3$ & $[a] \times [b] \times [a]$ & $n_1 m_2 n_3$ \\
$q_4$ & $[a] \times [b] \times [b]$ & $n_1 m_2 m_3$ \\
$q_5$ & $[b] \times [a] \times [a]$ & $m_1 n_2 n_3$ \\
$q_6$ & $[b] \times [a] \times [b]$ & $m_1 n_2 m_3$ \\
$q_7$ & $[b] \times [b] \times [a]$ & $m_1 m_2 n_3$ \\
$q_8$ & $[b] \times [b] \times [b]$ & $m_1 m_2 m_3$ \\
\hline
\end{tabular}\label{tab:cyclebasis}
\caption{The choice of basis for the 3-cycles}
\end{center}
\end{table}
The tadpole cancellation condition for an orientifold assumes the form,
\begin{eqnarray}
\sum{N_a (\vec{q}_a+\vec{q}_{a^{*}})}=4\vec{q}_{ori}.
\end{eqnarray}
The transformation can also be applied to the orbifolded models. For the $\mathbb{Z}_4 \times \mathbb{Z}_4$-orbifold only two of the eight vector entries are nonvanishing: The first entry, which contains the charge of the sector $g_0=g_1^{k_1} g_2^{k_1} g_3^{k_1}$ with $k_1=k_2=k_3=1$, and the last entry, with the charge of the twisted sector $g_0^3=g_1^{k_1} g_2^{k_1} g_3^{k_1}$ where $k_1=k_2=k_3=3$. Using the fact that all eight choices can also be expressed in terms of $g_0, h_1$ and $h_2$, we can use the same transformation to obtain the large-volume A- and B-side charges for the $\mathbbm{T}^6/\mathbb{Z}_2 \times \mathbb{Z}_2$. The only difference is that in the end result, the orbits have to be taken in the usual way, which is just an overall prefactor of $4$ for the $\mathbb{Z}_2 \times \mathbb{Z}_2$ group.
\\\\Let us extract the homology charges for a few examples.
\subsection{Example 1: Fundamental Cycles}
Tensoring together three branes of Eq.~(\ref{eq:T2branes}), computing their charge matrix and acting with $T^A T^{LV}$ on it, we obtain a matrix with entries 1 in the anti-diagonal and 0 everywhere else. These are the eight fundamental cycles spanning the basis shown in Table~\ref{tab:cyclebasis} and from the branes wrapping these cycles it is easy to obtain branes with any wrapping number by tachyon condensation.
\subsection{Example 2: Non-factorizable Cycles}
Let us derive the wrapping numbers for the $6\times 6$-factorizations on the $\mathbbm{T}^4$. There are three different types:
\begin{eqnarray}
\begin{array}{cc}
E_1=\begin{pmatrix}
 X_1 Y_1 & Y_1 Y_2 & X_1 X_2\\
 X_1 X_3 X_4  & -X_3 X_4 Y_2 & Y_2 Y_3 Y_4\\
 Y_1 Y_3 Y_4 & X_2 X_3 X_4 & -X_2 Y_3 Y_4
\end{pmatrix}\otimes (z_1)
&
J_1=\begin{pmatrix}
 0 & X_2 & Y_2 \\
 Y_3 Y_4 & 0 & X_1\\
 X_3 X_4 & Y_1 & 0 
\end{pmatrix}\otimes (-z_1),\\\\
E_2=\begin{pmatrix}
 X_1 Y_2 & Y_2 Y_3 Y_4 & X_1 X_2\\
 X_1 X_3 X_4  & -X_3 X_4 Y_3 Y_4 & Y_1 Y_3 Y_4\\
 Y_1 Y_2 & X_2 X_3 X_4 & -X_2 Y_1 \\
 \end{pmatrix}\otimes (z_1)
&
J_2=\begin{pmatrix}
 0 & X_2 & Y_3 Y_4\\
 Y_1 & 0 & X_1  \\
 X_3 X_4 & Y_2 & 0
 \end{pmatrix}\otimes (-z_1),\\\\
E_3=\begin{pmatrix}
-X_1 Y_3 Y_4 & Y_2 Y_3 Y_4 & X_1 X_2\\
 X_1 X_3 X_4  & -X_3 X_4 Y_2 & Y_1 Y_2\\
 Y_1 Y_3 Y_4 & X_2 X_3 X_4 & -X_2 Y_1 \\
\end{pmatrix}\otimes (z_1)
&
J_3=\begin{pmatrix}
 0 & X_2 & Y_2 \\
 Y_1 & 0 & X_1  \\
 X_3 X_4 & Y_3 Y_4 & 0  
\end{pmatrix}\otimes (-z_1).
\end{array}
\end{eqnarray}
Eq.~(\ref{eq:Rchargeformula}) defines a matrix with R-charges which can be transformed into the usual homology charges by acting on it with $T^A$ and $T^{LV}$. The result is,
\begin{eqnarray}
\begin{array}{l}
T^A T^{LV} \langle Q_1^{(n_1,n_2)} | k_1,k_2 \rangle
=\frac{1}{2}\begin{pmatrix}
 0 & 1 & 1 & 1\\
 1 & 0 & 1 &-1 \\
 1 & 1 & 0 &-1\\  
 1 &-1 &-1 & 0
\end{pmatrix},
\\\\
T^A T^{LV} \langle Q_2^{(n_1,n_2)} | k_1,k_2 \rangle
=\frac{1}{2}\begin{pmatrix}
-1 & 0 & 0 & 1\\
 0 & 1 & 1 & 0 \\
 0 & 1 & 1 & 0\\  
 1 & 0 & 0 &-1
\end{pmatrix},
\\\\
T^A T^{LV} \langle Q_3^{(n_1,n_2)} | k_1,k_2 \rangle
=\frac{1}{2}\begin{pmatrix}
 1 & 1 & 1 & 0\\
 1 &-1 & 0 &-1 \\
 1 & 0 &-1 &-1\\  
 0 &-1 &-1 & 1
\end{pmatrix}.
\end{array}
\end{eqnarray}
Note that in the notation used, the brackets on the left hand represent the entire charge matrix. These are fractional branes, thus the prefactor of $1/2$. Every column contains the wrapping numbers of one orbit of the brane. 
The last column for brane $Q_2$ for instance encodes the homology class $[(1,0)(1,0)+(0,1)(0,-1)]$. That particular brane had been constructed (non-topologically) from a coisotropic brane in~\cite{coisotropic}, section 7.2 and the defining equation of its physical locus in complex coordinates is $z^1=\bar{z}^2$.\\\\
If the first column represents the charge of a factorization $Q$ invariant under the pair of generators $\gamma_1\equiv \gamma(g_1)$ and $\gamma_2 \equiv \gamma(g_2)$, then the homology charges of the four columns correspond to factorizations invariant under $\alpha_1 \gamma_1$ and $\alpha_2 \gamma_2$ with the pair of phases $(\alpha_1,\alpha_2)$ given in this table:\\\\
\begin{tabular}{|c|cccc|}
\hline
column & 1 & 2 & 3 & 4\\
\hline
$(\alpha_1,\alpha_2)$& $(1,1)$ & $(i,1)$ & $(1,i)$ & $(i,i)$\\
\hline
\end{tabular}
\\\\
A similar correspondence holds for the eight columns of the $T^6$-charge matrix. Here, a brane is invariant under three generators $\alpha_1 \gamma_1$, $\alpha_2 \gamma_2$ and $\alpha_3 \gamma_3$ with the following phases $(\alpha_1,\alpha_2,\alpha_3)$:\\\\
\begin{tabular}{|c|cccccccc|}
\hline
column & 1 & 2 & 3 & 4 & 5 & 6 & 7 & 8\\
\hline
$(\alpha_1,\alpha_2,\alpha_3)$ & $(1,1,1)$ & $(i,1,1)$ & $(1,i,1)$ & $(i,i,1)$
&$(1,1,i)$ & $(i,1,i)$ & $(1,i,i)$ & $(i,i,i)$\\
\hline
\end{tabular}
\\\\This is in line with our convention of the eight basis elements of the $\vec{q}$-basis.
\subsection{Tachyon Condensation}
\begin{figure}
\begin{center}
\scalebox{1.2}{\includegraphics{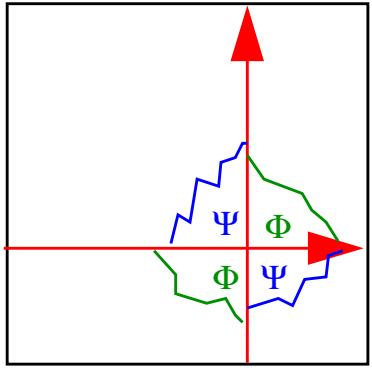}}
\end{center}
\caption{Drawn are two branes with wrapping numbers (1,0) and (0,1) on the A-side. A boson or a fermion are located at the intersection, depending on
the orientation of the brane, which may not be know a priory. Brane and anti-brane described by the same matrix factorization in this model. Since we know that the wrapping numbers add up when the fermion $\Psi$ enters in the bound state formation process, we know it is located in the corner where it is shown. For the boson $\Phi$, the resulting homology charge is the difference between the branes, so in that case one of the arrows would have to be reversed.}
\label{fig:tachcond}
\end{figure}

It is well known that two factorizations with a morphism between them can form a bound state and give rise to a new
factorization. The new factorization is obtained from the cone construction (see e.g.~\cite{tachcubic}). Here we will
briefly analyze how the new orbifold generators are obtained and how they give rise to the homology class of the
bound state.\\
Suppose there are two branes on the $\mathbb{T}^2$ denoted by $(Q_a,\gamma^a_1,\gamma^a_2,\gamma^a_3)$ and $(Q_b,\gamma^b_1,\gamma^b_2,\gamma^b_3)$. Morphisms $\Phi^{ab}(x_i), \Psi^{ab}(x_i) \in \mbox{Hom}(Q_a,Q_b)$ stretching between the two branes
must satisfy the orbifold invariance condition:
\begin{eqnarray}
\gamma^b_i\; \Psi^{ab}(g_i x_i)\; (\gamma^a_i)^{-1}=\Psi^{ab}(x_i)\qquad \mbox{for }i=1,2,3.
\end{eqnarray}
The other orbifold models are subject to analogous conditions.
For later convenience, we write an orbifold generator $\gamma$ on the $\mathbb{Z}_2$-graded space in block-diagonal form:
\begin{eqnarray}
\gamma=\mbox{diag }(\gamma_{+},\gamma_{-})=\begin{pmatrix}\gamma_{+} & 0\\ 0 & \gamma_{-} \end{pmatrix},
\end{eqnarray}
so that the supertrace reads,
\begin{eqnarray}
\mbox{STr }\gamma=\mbox{Tr }\gamma_{+} - \mbox{Tr }\gamma_{-}.
\end{eqnarray}
Bound state formation with a boson given by,
\begin{eqnarray}
\Phi^{ab}=\begin{pmatrix}\phi_0^{ab} & 0\\ 0 & \phi_1^{ab} \end{pmatrix},
\end{eqnarray}
or with a fermion denoted by,
\begin{eqnarray}
\Psi^{ab}=\begin{pmatrix} 0 & \psi_0^{ab}\\ \psi_1^{ab} & 0 \end{pmatrix},
\end{eqnarray}
results in the following bound state factorizations:
\begin{eqnarray}
F_{bos}^{ab}=\begin{pmatrix} F^a & -\phi_0^{ab} \\ 0 & G^b\end{pmatrix}\qquad
G_{bos}^{ab}=\begin{pmatrix} G^a & \phi_1^{ab} \\ 0 & F^b\end{pmatrix}.\label{eq:tachcon1}
\end{eqnarray}
and,
\begin{eqnarray}
F_{fer}^{ab}=\begin{pmatrix} F^a & \psi_0^{ab} \\ 0 & F^b \end{pmatrix}\qquad
G_{fer}^{ab}=\begin{pmatrix} G^a & \psi_1^{ab} \\ 0 & G^b\end{pmatrix}.\label{eq:tachcon2}
\end{eqnarray}
It is easy to see that the orbifold generator for the new branes are,
\begin{eqnarray}
\begin{array}{rcl}
\displaystyle\gamma_{bos}&=&\displaystyle\mbox{diag }(\gamma^a_{+},\gamma^b_{-},\gamma^a_{-},\gamma^b_{+}),\\\\
\displaystyle\gamma_{fer}&=&\displaystyle\mbox{diag }(\gamma^a_{+},\gamma^b_{+},\gamma^a_{-},\gamma^b_{-}),
\end{array}
\end{eqnarray}
where each entry in the diagonal matrix is a block-matrix once again.
Consequently, for the supertrace holds,
\begin{eqnarray}
\begin{array}{rcl}
\mbox{STr }(\gamma^{bos})^k &=&\mbox{STr }(\gamma^{a})^k-\mbox{STr }(\gamma^b)^k,\\
\mbox{STr }(\gamma^{fer})^k &=&\mbox{STr }(\gamma^{a})^k+\mbox{STr }(\gamma^b)^k.
\end{array}
\end{eqnarray}
For a product of several generators in the supertrace the formula is still valid, so that together with Eq.~(\ref{eq:Rchargeformula})
we know that after tachyon condensation with a boson (or fermion) the resulting charge matrix is a difference (or sum) of the charge matrices
of the original branes. Of course, from the linearity of $T^{LV}$ and $T^{A}$ the homology charge matrices subtract or add in the same manner as expected.\\
We can use these results to locate the position of the fermion and boson at the intersection. It is not necessarily clear in which of the four corners of the intersection the boson is found and where the fermion. By using the derived results, we can visualize the location of the boson and fermion in the intersection as is shown exemplary in Fig.~\ref{fig:tachcond}. The issue arises because the factorizations a priori do no distinguish
branes from anti-branes, so there will be a bosonic as well as a fermionic state. One of them is present, depending on what are branes and what anti-branes. Would we discuss a LG-theory with odd degree like a cubic curve, we would not have to deal with such an ambiguity since there branes and anti-branes are given by disctinct factorizations.\\\\
Of course, the addition of these K-theory charges always takes place no matter if the branes annihilate, if the combined brane is instable and decays into two or more new D-branes or if a stable new bound state appears. It should be noted that the cases are distinguished easily in this framework: The branes annihilate if the new factorization is isomorphic to the trivial factorization $\mathbbm{1}W=W\mathbbm{1}=W$, the branes decay into new branes if the factorization is a direct sum of lower-dimensional factorization and the branes form a new bound-state if neither of this happens.\\
We are now able in a position to obtain {\it any} possible BPS-brane in the discussed background: The factorization $Q_a$ and its orbits span a basis of the charge lattice. By an appropriate sequence of tachyon condensations we get a brane with arbitrary homology class. The resulting brane can then be shifted away from the fixed points as described in section~\ref{sec:moduli}. Some stability questions have been discussed in~\cite{harunT6}.\\
A similar construction should be possible for other Calabi-Yau manifolds. Start with a minimal set of branes (and anti-branes) with charge vectors $q_1,...,q_N$. For every homology charge
$\mathbb{N}q_1+...+\mathbb{N} q_N$ a brane can be found. The modulus of this brane can then be continously changed so as to move it to the desired location and (on the A-side) to turn on the desired Wilson line. Of course, branes incompatible with the orbifold action must be projected out.
To every brane which is not invariant under the orientifold action its orientifold image brane has to be added as well. The orientifold action is explored further below in section~\ref{sec:orientifold}. Finding the set of branes to start with should not be hard provided that flat coordinates
have been found like in section~\ref{sec:moduli}. After all, the charges of such branes shall be as small
as possible and a rough correspondence between the size of the factorizations and the magnitude of its charge exists. The smaller the charge, the smaller the factorization. The branes of the starting set are most likely the simplest factorizations. Namely tensor products of the so-called permutation branes which describe the fractional branes. It is a priori not clear if matrix factorizations can describe all existing branes. The construction proves that this is indeed the case for the considered orbifolds and provides a way to test this for other backgrounds.
\section{Orientifold-Planes}
\label{sec:orientifold}
What kind of orientifold actions compatible with the background exist? What is the homology charge of their orientifold planes? How does the orientifolding act on a D-brane?\\
In the LG framework, the orientifold action arises as an involution $\sigma$ acting on the LG potential $W(x_i) \rightarrow W(x_i)$. The advantage is that all possible orientifolds can be found easily and not just
by guessing which ones could be compatible with the background geometry. $\textbf{A}$- and $\textbf{B}$-type orientifold lattices, orientifold that act  on Wilson lines trivially or with a discrete shift, and orientifolds that act on the position moduli trivially or with a discrete shift can all be found. Here, the analysis of~\cite{harunT6} is extended and the $\mathbbm{T}^6/\mathbbm{Z}_2 \times \mathbbm{Z}_2$ is analyzed.\\\\
The conditions for orientifold invariance on a brane $Q(x_i)$ in a quantum orbit $\gamma(g)$ 
are~\cite{orientifoldcat},
\begin{eqnarray}
\begin{array}{l}
\displaystyle U(\sigma)Q^T(\sigma x_i)U(\sigma)^{-1}=Q,\\\\
\displaystyle U(\sigma)(\chi(g) \gamma^{-T}(g))U(\sigma)^{-1}=\gamma(g).\label{eq:invarcond}
\end{array}
\end{eqnarray} 
The orientifold action is defined by an involution $\sigma$ together with a character $\chi: \Gamma \mapsto \mathbb{C}$.
The first line of Eq.~(\ref{eq:invarcond}) purports that the brane is invariant under the involution $\sigma$ if a transformation $U$ on the $\mathbb{Z}_2$-graded space can be found such that after the orintifold action the factorization is isomporphic to the original factorization.
The second line of the equation keeps track of the quantum orbit the orientifold image is mapped into -- not only the factorization has to stay unchanged, the orbit of the brane must remain invariant as well. The choice of a phase $\chi$ selects the desired value of that orbit.\\
We now want to compute the homology charges for the different orientifold involutions. On each $\mathbbm{T}^2$ the full set of possible orientifold parities is given by~\cite{harunT6},
\begin{eqnarray}
\sigma_1^{(n,m)}: (x_1,x_2,z_1) \mapsto (e^{n\pi i(\frac{1}{4}+\frac{n}{2})}x_1,e^{n\pi i(\frac{1}{4}+\frac{m}{2})} x_2,i z_1),\;\;n+m=0\mbox{ mod }2,\\
\sigma_2^{(n,m)}: (x_1,x_2,z_1) \mapsto (e^{n\pi i(\frac{1}{4}+\frac{n}{2})}x_2,e^{n\pi i(\frac{1}{4}+\frac{m}{2})} x_1,i z_1),\;\;n+m=0\mbox{ mod }2.\label{eq:invol}
\end{eqnarray}
Up to an overall sign, the R-charges of the crosscap states can be obtained from a formula derived in~\cite{orientifoldcat}. For trivial character $\chi=1$ the resulting charges are, 
\begin{eqnarray}
\begin{array}{rl}
O_1^{(0,0)}:& T^A\; T^{LV}\; ( 4i,-4i)^T=(1,1)^T,\\
O_2^{(0,0)}:& T^A\; T^{LV}\; ( 4,4)^T=(-1,1)^T.
\end{array}
\end{eqnarray}
The O-planes corresponding to these involutions stretch across the torus diagonal. They are the so-called $\textbf{B}$-type involutions as opposed to the standard $\textbf{A}$-type involution $\tilde O_1$ shown further below whose O-planes lie parallel to a torus axis. In the model-building literature this action is well-known, but here it looks unusual because the lattice of the wrapping number is rotated: When the $\textbf{B}$-type lattice is used, the axis on which the branes are reflected is usually taken to be the $x$-axis and torus lattice vectors are taken to be $(1,1)$ and $(1,-1)$. That is, the entire torus is rotated and the orientifold acts according to $(n,m)\rightarrow (n,-m)$ on the wrapping numbers in both the $\textbf{A}$- and the $\textbf{B}$-lattice. In our Landau-Ginzburg description, on the other hand, the torus lattice remains fixed and the orientifold lattice can be embedded in different ways. For the $\textbf{A}$-lattice it acts as before by inverting the second wrapping number. Other choices of the involution can also invert $n$ instead of $m$. In the $\textbf{B}$-lattice the orientifold action maps $(n,m)\rightarrow (m,n)$.
In terms of an action in complex coordinates on the target space, this would amount to modifying the usual orientifold action $z \mapsto \bar{z}$ by multiplying it with the imaginary unit,
\begin{eqnarray}
\mathcal{R}: z \mapsto i\bar{z}.
\end{eqnarray}
In the LG model discussed here, the more common $\textbf{A}$-type lattice is obtained by setting the character to $\chi=i$. The charges are then,
\begin{eqnarray}
\begin{array}{rl}
\tilde O_1^{(0,0)}:& T^A\; T^{LV}\; (-4+4i,-4-4i)^T=(-2,0)^T,\\
\tilde O_2^{(0,0)}:& T^A\; T^{LV}\; (0,0)^T=(0,0)^T.
\end{array}
\end{eqnarray}
A $\tilde O_1^{(0,0)}$ action on all three tori of a plain $\mathbbm{T}^6$ for example is the most commonly used orientifold and the total homology charge of the O-planes is,
\begin{eqnarray}
[(-2,0)(-2,0)(-2,0)]=-8[(1,0)(1,0)(1,0)].
\end{eqnarray}
An further prefactor of 4 is necessary to take into account the space-time contribution. This is the charge used in the tadpole cancellation condition. Different choices for $n,m$ result in homology charges which are equivalent up to rotations of the O-planes by multiples of $\pi/2$ on the tori.\\\\
The $\mathbbm{T}^6 /\mathbb{Z}_2 \times \mathbb{Z}_2$ LG-orbifold has three quantum generators which were denoted $g_0$, $h_1$ and $h_2$. The single-generator formulas in~\cite{orientifoldcat} generalize in a straight-forward manner to multiple-generator theories. There can however be different characters for every generator. For consistency I impose,
\begin{eqnarray}
\chi(1)=1 \qquad \chi(g_a)\chi(g_b)=\chi(g_a g_b) \;\;\forall \;\; g_a,g_b \in \Gamma.
\end{eqnarray}
The only possible choices for the characters of the group generators would therefore be $\chi(g_0)=\pm i$ and $\chi(h_{1,2})=\pm 1$.\\
In~\cite{harunT6} a type of orientifold plane was found which has apparently not been analyzed to date despite the ubiquity of the $\mathbbm{T}^6 /\mathbb{Z}_2 \times \mathbb{Z}_2$ orientifolds in the literature.\\
Namely, the orientifold action can have a discrete $\mathbb{Z}_2$-shift on position moduli (and Wilson line moduli) on top of the inversion.
In the A-side picture of square tori, such an orientifold has different lattices for the torus, for the orbifold action as well as for the orientifold action. The O-plane neither has to concide with the torus axis, nor does it have to pass through orbifold fixed points. The different actions on the $\mathbbm{T}^2$ building blocks are recapped from~\cite{harunT6} in a few lines, thereafter results for the $\mathbbm{T}^6 /\mathbb{Z}_2 \times \mathbb{Z}_2$ orientifold are derived from it.\\
\begin{figure}
\begin{center}
\scalebox{0.6}{\includegraphics{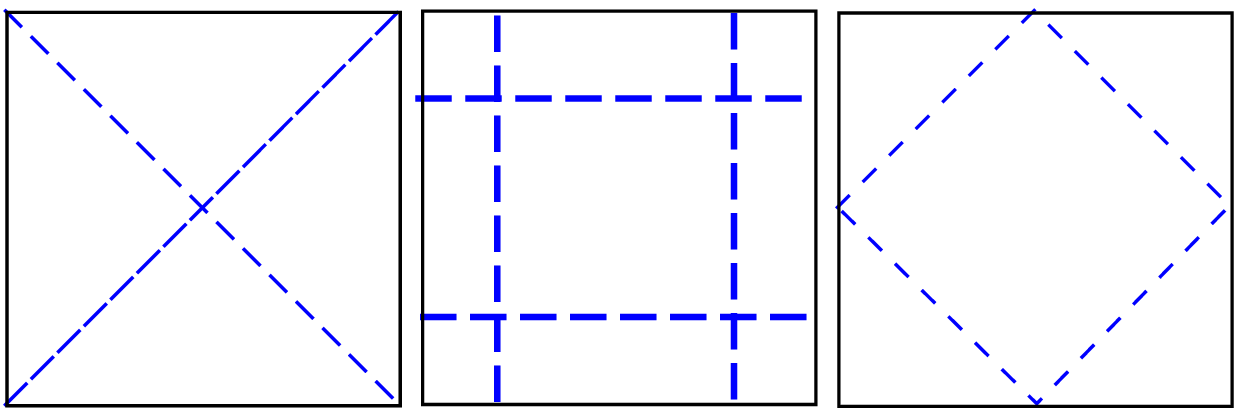}}
\scalebox{0.6}{\includegraphics{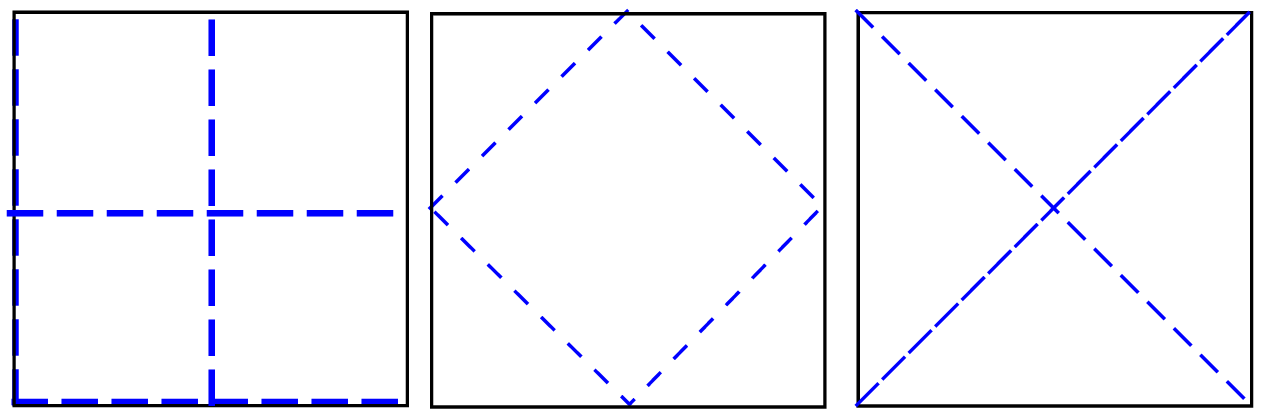}}
\end{center}
\caption{Two different examples of orientifold planes in $\mathbbm{T}^2 \times \mathbbm{T}^2 \times \mathbbm{T}^2 /\mathbb{Z}_2 \times \mathbb{Z}_2$ are shown. Note that the locations of the orientifold planes, the torus lattice vectors and orbifold fixed points do not necessarily coincide. Instead, the orbifold action can exchange different O-planes. The orientifold actions for O-planes passing not through the origin have a $\mathbb{Z}_2$-shift on the coordinate. ($\mathbb{Z}_2$-shifts on the Wilson line are also possible.) These involutions were analyzed in~\cite{harunT6}.}\label{fig:Oplanes}
\end{figure}
By studying the effects of the orientifold action on the moduli of fundamental cycles on the torus it was found that $\sigma_1$ acts trivially on the Wilson line component in the corresponding torus whereas $\sigma_2$ incorporates an additional $\mathbb{Z}_2$-shift in the Wilson line. Furthermore, it could be seen that $\sigma_i^{(n+1,m+1)}$ and $\sigma_i^{(n,m)}$ differ only by a rotation of the orientifold plane by $\pi/2$. (This was established as follows: The modulus picks up a sign for one involution but not for the other; since the modulus contains the distance to the lattice vector on the A-side, a sign flip indicates a reflection of the test cycle by the lattice vector, therefore it must be parallel to the O-plane. If there is no sign change, the brane is left invariant, which is the case for a brane orthogonal to an O-plane.) This can of course also be seen from the homology charge of the orientifold plane or by realizing that $\sigma_{1,2}^{(n+1,m+1)}=i \sigma_{1,2}^{(n,m)}$ differs from $\sigma_{1,2}^{(n,m)}$ by the phase $i$ which is nothing but a shift in the quantum orbit generated by the corresponding $g_j$, $j=1,2,3$. Therefore $g_j \sigma_{1,2}$ rotates the O-plane on the $j$-th torus. Finally, for $\sigma_1^{(n,m)}$ with $n\ne m$ or $\sigma_2^{(n,m)}$ with $n=m$, the action includes an additional $\mathbb{Z}_2$-shift of the modulus and the O-plane is shifted away from the origin. The modulus $u$ parametrizing a shift from the origin is mapped into the modulus $u^{*}=-u+\frac{1}{2}$ (plus a possible action on the Wilson line). An invariant brane would have $u^{*}=u \mbox{ mod } 1$ where mod 1 is due to the identification of points by the torus. Invariance is therefore obtained for $u=\frac{1}{4}$ or $\frac{3}{4}$.\\
As noted at the beginning of the section, invariance with respect to the orientifold action must not only signify invariance of the modulus, it also means that the orbit of the brane is preserved. That requirement is encoded in the second line of Eq.~(\ref{eq:invarcond}). If for some characters $\chi$ and certain moduli the fundamental cycles $(1,0)$ or $(0,1)$ are invariant, the O-planes lie parallel to a torus lattice vector. If, on the other hand, the diagonals $(1,1)$ and $(1,-1)$ are mapped into themselves, it means that the O-planes lie parallel to the diagonals. The lattices obtained for different characters are compiled in the following table:
\begin{center}
\begin{tabular}{|c|cccccccc|}
\hline
& \textbf{AAA} & \textbf{AAB} & \textbf{ABA} & \textbf{BAA} & \textbf{ABB} & \textbf{BAB} & \textbf{BBA} & \textbf{BBB} \\
\hline
$\chi(g_0)$ & $i$ & 1 & 1 & 1 & $i$ & $i$ & $i$ & 1\\
$\chi(h_1)$ & 1 & 1 & -1 & -1 & -1 & -1 & 1 & 1\\
$\chi(h_2)$ & 1 & -1 & -1 & 1 & 1 & -1 & -1 & 1\\
\hline
\end{tabular}
\end{center}
The table was constructed by probing the invariance of a test cycle. The results are consistent with the expected homology charges. The smallest homology charge has the \textbf{BBB}-model since every \textbf{B}-lattice halves the number of orientifold planes and their greater charge (they wrap a diagonal and not only a fundamental cycle) can not fully make up for the reduction. The homology charges of the different types of lattices are,
\begin{eqnarray}
\begin{array}{rcl}
\textbf{AAA}&:&(8,0,0,-8,0,-8,-8,0)^T\\
\textbf{AAB}&:&(4,-4,-4,-4,-4,-4,-4,4)^T\\
\textbf{ABB}&:&(4,0,0,-4,0,-4,-4,0)^T\\
\textbf{BBB}&:&(2,-2,-2,-2,-2,-2,-2,2)^T
\end{array}
\end{eqnarray}
These charges were derived from the involution $\sigma_1^{(0,0)} \times \sigma_1^{(0,0)} \times \sigma_1^{(0,0)}$.
Rotations of planes on the tori lead to permutations of the vector entries. Other involutions can give rise to either sign reversals in some vector entries or to the complete vanishing of $\vec{q}_{ori}$.\\
Let us quickly demonstrate the reversal of the process, that is, how the LG-involution for a certain large volume O-plane configuration can be obtained. Two examples are illustrated in Fig.~\ref{fig:Oplanes}. The first is a \textbf{BAB} lattice and the second one an \textbf{ABB} lattice. The characters can be looked up in the above table. For the branes shifted
away from the origin, we need $n \ne m$ in $\sigma_1$ of Eq.~(\ref{eq:invol}) of the corresponding torus. The corresponding
involutions are therefore $\sigma_1^{(0,0)} \times \sigma_1^{(0,2)} \times \sigma_1^{(0,2)}$ and $\sigma_1^{(0,0)} \times \sigma_1^{(0,2)} \times \sigma_1^{(0,0)}$. Again, $(n,m)\mapsto (n+1,m+1)$ on some torus would merely rotate the O-plane. A $\mathbb{Z}_2$-action on the Wilson line could be modeled with $\sigma_2$. See~\cite{harunT6}. 
\section{Gauge Groups and Higgsing}
A stack of $N$ D-branes in generic position supports a $U(N)$ gauge group on it. Splitting a stack of orientifold invariant branes and moving branes away from the O-plane is the D-brane analogue to the field-theoretic Higgs mechanism. On the face of it this sounds straightforward but things become
a bit more tricky when an orbifold action is present. The D-brane can be in the bulk, or it can go through fixed points and be orbifold invariant. A brane with homology charge identical to an O-plane can be shifted away from the symmetry axes in one, two or three tori and it can happen that each cycle
is orientifold invariant, that cycles are invariant pairwise or that they are not invariant at all. When the branes are non-factorizable, these questions are harder to answer since it is harder to visualize them. For a brane on a generic Calabi-Yau it is also not obvious to find the gauge group -- whether unitary, symplectic or orthogonal. For fundamental cycles on the $\mathbbm{T}^6 / \mathbb{Z}_2 \times \mathbb{Z}_2$, all these
questions have been answered in~\cite{D6branesplitting}. The purpose of this section is to demonstrate that the framework employed in this paper is well capable of answering these questions by showing how the results of the cited paper can be reproduced.\\\\
In the $\mathbbm{T}^6 / \mathbb{Z}_2 \times \mathbb{Z}_2$-orbifold we can chose the standard orientifold involution $\sigma=\sigma_1^{(0,0)}\otimes \sigma_1^{(0,0)} \otimes \sigma_1^{(0,0)}$ and $\chi(g_0)=-i$. The sum of the homology charge of the O-planes is $q(\sigma)=(2,0,0,-2,0,-2,-2,0)^T$.
Let us place a D-brane right on top of an orientifold plane and ask: What happens to the gauge group if we move the brane away from the O-plane in one, two or three tori?\\
The invariance conditions on a brane $Q$ in a quantum orbit $\gamma(g)$ were given in Eq.~(\ref{eq:invarcond}).
A brane on top of one of the orientifold planes is the brane discussed extensively in~\cite{harunT6}, which was schematically denoted by,
\begin{eqnarray}
Q_4 \equiv Q^{(2)} \otimes Q^{(2)} \otimes Q^{(2)} \otimes
\begin{pmatrix}
0 & -z_1\\z_1 & 0
\end{pmatrix}.
\end{eqnarray}
A similarity transformation $U_c(\sigma)$ for $Q_c$ satisfying Eq.~(\ref{eq:invarcond}) can be found easily, but due to lack of space I do not write out the $16\times 16$ matrices here. The brane can be shifted away from the fixed points in one of the tori,
\begin{eqnarray}
Q_5 \equiv Q^{(2)} \otimes Q^{(2)} \otimes Q^{(2)} \otimes
\begin{pmatrix}
0 & -z_1+d_1 x_1 x_2\\z_1+d_1 x_1 x_2 & 0
\end{pmatrix},
\end{eqnarray}
This modification adds $d_1^2 x_1^2 x_2^2$ to the superpotential which moves the brane into the bulk.
The orientifold invariance Eq.~(\ref{eq:invarcond}) still holds. This means the gauge group remains symplectic.
A stack of two such branes can be split and moved into the bulk in the remaining two tori. The factorization of such a brane is,
\begin{eqnarray}
\begin{array}{l}
E_6=
\begin{pmatrix}
E_5 & (d_2 x_3 x_3+i d_3 x_5 x_6)\mathbbm{1}\\
 -(d_2 x_3 x_4-i d_3 x_5 x_6)\mathbbm{1} & -J_5
\end{pmatrix},\\
J_6=
\begin{pmatrix}
J_5 & -(d_2 x_3 x_4+i d_3 x_5 x_6)\mathbbm{1}\\
 (d_2 x_3 x_4-i d_3 x_5 x_6)\mathbbm{1} & E_5
\end{pmatrix},\\
Q_6=\begin{pmatrix}
0 & E_6\\
J_6 & 0
\end{pmatrix}.
\end{array}
\end{eqnarray}
Note that the factorization is not symmetric under exchange of $d_1 x_1 x_2$ with $d_2 x_3 x_4$ or $d_3 x_5 x_6$. But this is only apparent; under the swapping the branes remain isomorphic as can be shown easily by finding a similarity transformation relating the factorization. By adding two additional quadratic terms to the potential the isomorphism becomes manifest: Then the brane can be written as a tensor product of the factorization $E_2$, $J_2$ and three factorizations of the type $(z_1+d_1 x_1 x_2)(-z_1+d_1 x_1 x_2)$.\\
Clearly the brane $Q_6$ decomposes into a direct sum of two identical branes for $d_2=d_3=0$. A similarity transformation compatible with Eq.~(\ref{eq:invarcond}) does not exist for generic values of $d_2$ and $d_3$. Consequently, the brane is not invariant under the orientifold action and therefore supports a $U(1)$ gauge group. (A direct sum of $N$ branes as ususal gives rise to a $U(N)$ gauge group.) If either $d_2$ or $d_3$ is set to zero and the other modulus takes a generic value, however, the brane is orientifold invariant. But since the direct sum of two identical branes is broken to a single brane by turning on the modulus, the degree of the gauge group is halved. In short, we can determing whether the gauge group of a stack of branes is unitary or not by checking whether a similarity transformation compatible with Eq.~(\ref{eq:invarcond}) exists. The degree of the gauge group is directly linked to the number of identical factorizations into which the brane can be decomposed as a direct sum.
This is a relatively easy way to determine the gauge group and the results agree perfectly with those obtained for the selected example in~\cite{D6branesplitting}. This method works for general Calabi-Yau in likewise manner.  In~\cite{orientifoldcat} it is described how a symplectic gauge group can be distinguished from an orthogonal group in the formalism used here.
\section{The physical locus of a D-brane}
\label{sec:loc}
A topological D-brane is specified by its homology class and its moduli. 
It is however not always easy to see how a D-brane of given homology class
lies in space (up to shifting it around by virtue of the position moduli, of course).
The basic T-duality formulas -- relating the fields of a magnetized D9-brane
on the IIB side with a D6-brane lying somewhere in space on the IIA-side mirror --
are well-known but once again the computation seems
not to have been performed except in the simplest of cases. 
Since this work deals with non-factorizable branes on the torus
and we would like to have a geometrical interpretation of such branes
it seems proper to fill the gap and perform the analysis.
While it does not contain any new piece of physical information of the low-energy
theory, the ability to visualize D-branes is certainly a very attractive
feature and can improve the intuitive understanding.
Furthermore, the connection to magnetized branes on the IIB-side is necessary
for the identification of the coisotropic branes. Coisotropic branes correspond
to branes $a$ and their image brane $b$ with $[F_a,F_b]\ne 0$. 
At the level of matrix factorizations no way is known to distinguish coisotropic branes
from other branes.
We now want to determine the geometrical locus of a D-brane on the IIA side
for given magnetic fields on the worldvolume of the IIB side brane.
This is done in full generality for the $\mathbbm{T}^4$.
\subsection{Cayley-transformation}
A D-brane with a general constant magnetic field on its world-volume has the boundary condition,
\begin{eqnarray}
\partial_{\sigma} X_i -2 \pi \alpha' F_{ij} \partial_{\tau} X^j=0,\qquad \sigma=0,\pi.
\end{eqnarray}
Rewritten in the light-cone frame,
\begin{eqnarray}
\partial_{+}=\frac{1}{2}(\partial_{\tau}+\partial_{\sigma})\qquad
\partial_{-}=\frac{1}{2}(\partial_{\tau}-\partial_{\sigma}),
\end{eqnarray}
the boundary condition becomes,
\begin{eqnarray}
(\delta_{ij}-2 \pi \alpha' F_{ij}) \partial_{+} X^j=(\delta_{ij}+ 2 \pi \alpha' F_{ij}) \partial_{-} X^j,
\end{eqnarray}
or,
\begin{eqnarray}
\partial_{+} X_i=A_{ij} \partial_{-} X^j, \label{eq:boundarycond}
\end{eqnarray}
where,
\begin{eqnarray}
A=(I-2 \pi \alpha' F)^{-1} (I+ 2 \pi \alpha' F).\label{eq:cayleytransform}
\end{eqnarray}
The matrix $A \in SO(N)$ is the so-called Cayley transform of $2 \pi \alpha' F$. The Cayley transform is its own inverse,
so from any orthogonal matrix $A$, the corresponding anti-symmetric tensor $2 \pi \alpha' F$ can be obtained by the same
formula with $A$ and $2 \pi \alpha' F$ replaced.
\subsection{T-Duality in $\mathbbm{T}^4$ with general magnetic fields}
In analogy to the well-known $\mathbbm{T}^2$ case analyzed for example in~\cite{zwiebach}, consider a brane which wraps two dimensions on a $\mathbbm{T}^4$. In a frame in which the brane fills the $x'^2-x'^4$ plane, the two Dirichlet and two Neumann boundary conditions reduce to,
\begin{eqnarray}
\partial_{+}
\begin{pmatrix}
X'^1\\\tilde X'^2\\X'^3\\\tilde X'^4
\end{pmatrix}
=
\begin{pmatrix}
1 &0 & 0 & 0\\
0 &-1& 0 & 0\\
0 &0 & 1 & 0\\
0 &0 & 0 &-1
\end{pmatrix}
\partial_{-}
\begin{pmatrix}
X'^1\\\tilde X'^2\\X'^3\\\tilde X'^4
\end{pmatrix}.
\end{eqnarray}
Note that both the capital $X^i$ and the lower-case $x^i$ have different meanings here than in the preceeding sections. Here they are simply coordinates.
By acting on the equation with a matrix $R\in SO(4)$ we rotate the primed coordinates, which are aligned with the brane, into an unprimed frame in which a D2-brane T-dual to an unmagnetized D4-brane lies in the $x^2-x^4$-plane:
\begin{eqnarray}
\partial_{+}
\begin{pmatrix}
X^1\\\tilde X^2\\X^3\\\tilde X^4
\end{pmatrix}
=
R^{-1}
\begin{pmatrix}
1 &0 & 0 & 0\\
0 &-1& 0 & 0\\
0 &0 & 1 & 0\\
0 &0 & 0 &-1
\end{pmatrix}
R\; \partial_{-}
\begin{pmatrix}
X^1\\\tilde X^2\\X^3\\\tilde X^4
\end{pmatrix}.
\end{eqnarray}
By $T$-dualizing along the $\tilde x^2$ and $\tilde x^4$ directions, one obtains a space-filling brane with the Neumann boundary conditions:
\begin{eqnarray}
\partial_{+}
\begin{pmatrix}
X^1\\X^2\\X^3\\X^4
\end{pmatrix}
=
R^{-1}
\begin{pmatrix}
1 &0 & 0 & 0\\
0 &-1& 0 & 0\\
0 &0 & 1 & 0\\
0 &0 & 0 &-1
\end{pmatrix}
R\;\begin{pmatrix}
1 &0 & 0 & 0\\
0 &-1& 0 & 0\\
0 &0 & 1 & 0\\
0 &0 & 0 &-1
\end{pmatrix}
\partial_{-}
\begin{pmatrix}
X^1\\X^2\\X^3\\X^4
\end{pmatrix}.\label{eq:rotcomp}
\end{eqnarray}
This transformation must be identical to the operator $A$ of the Cayley-transform Eq.~(\ref{eq:cayleytransform}). The identification is the $T$-duality map in four dimensions with a general magnetic field on it: To a given field $F$ on the world-volume of a space-filling brane a rotation matrix $R$ is associated which encodes the tilted of the $T$-dualized brane with respect to the $T$-dual of a field free brane. Alternatively start with a tilted brane with a tilt defined by $R$ and derive the corresponding field $F$ on the dual side.\\
In order to establish the explicit mapping between the components of $F$ and the rotation parameters
it is convenient to switch to quaternion notation. By defining,
\begin{eqnarray}
X=X^1 + X^2 i + X^3 j+ X^4 k,
\end{eqnarray}
and decomposing the rotation into a left-isoclinic and a right-isoclinic factor,
\begin{eqnarray}
R_L=a+bi+cj+dk \qquad R_R=p+qi+rj+sk,
\end{eqnarray} 
the boundary condition can be rewritten in terms of quaternions. The quaternion basis obeys the algebra,
\begin{eqnarray}
i^2=j^2=k^2=ijk=-1.
\end{eqnarray}
To describe a rotation, the equivalent of an orthogonality condition must be fulfilled, which corresponds to the normalization of the quaternion:
\begin{eqnarray}
R_L^2\;=\;R_R^2\;=\;a^2+b^2+c^2+d^2\;=\;p^2+q^2+r^2+s^2\;=\;1.\label{eq:quatnorm}
\end{eqnarray}
In quaternion notation, the matrix $\mbox{diag}(1,-1,1,-1)$ simply
corresponds to the right- and left-isoclinic factors $j$ and $-j$. The boundary condition Eq.~(\ref{eq:rotcomp}) therefore
translates to,
\begin{eqnarray}
\partial_{+} X=\displaystyle  A_L(\partial_{-}X)A_R. 
\end{eqnarray}
with,
\begin{eqnarray}
\begin{array}{rcl}
\displaystyle A_L&=&\displaystyle (a-bi-cj-dk)(-1j)(a+bi+cj+dk)(-1j)\\
&=&\displaystyle a^2-b^2+c^2-d^2 -2(ab-cd)i+0j-2(bc+ad)k,\\\\
\displaystyle  A_R&=&\displaystyle  (p+qi+rj+sk)(1j)(p-qi-rj-sk)(1j)\\
&=&\displaystyle p^2-q^2+r^2-s^2 - 2(pq+rs)i+0j+2(qr-ps)k.
\end{array}\label{eq:Cayley1}
\end{eqnarray}
\subsection{T-Duality in 3D}
Before turning to the 4D case in the next section, consider a 3-dimensional subspace in this section for illustration.
Take a magnetized brane filling a 3-dimensional cube and $T$-dualize in two of the three directions.
To do this, disregard the first dimension in the above equations and turn on the most general field configuration in the remaining directions:
\begin{eqnarray}
F=\begin{pmatrix}
0 & 0 & 0 & 0\\
0 & 0 &F_{23} & F_{24}\\
0 &-F_{23} & 0 & F_{34}\\
0 & -F_{24} & -F_{34} & 0
\end{pmatrix}.
\end{eqnarray}
The Cayley-transform maps the field strenght tensor to the quaternion,
\begin{eqnarray}
\begin{array}{rcl}
A_L&=&\displaystyle\frac{1}{\sqrt{1+F_{23}^2+F_{24}^2+F_{34}^2}}(1-F_{34}i+F_{24}j-F_{23} k),\\\\
A_R&=&\displaystyle\frac{1}{\sqrt{1+F_{23}^2+F_{24}^2+F_{34}^2}}(1+F_{34}i-F_{24}j+F_{23} k),
\end{array}\label{eq:Cayley2}
\end{eqnarray}
Eqs.~(\ref{eq:Cayley1}) and~(\ref{eq:Cayley2}) match only if $F_{24}=0$. This reflects the fact that this
field is on the $x_2-x_4$ plane of the world-sheet which is $T$-dualized away. For three-dimensional rotations $A_L\equiv A_R$ always holds, so in components one gets,
\begin{eqnarray}
p=a \qquad q=-b \qquad r=-c \qquad s=-d.
\end{eqnarray}
Therefore, the $T$-duality map in 3 dimensions is:
\begin{eqnarray}
\begin{array}{l}
\displaystyle \frac{1}{\sqrt{1+F_{23}^2+F_{34}^2}}=a^2-b^2+c^2-d^2,\\
\displaystyle \frac{F_{34}}{\sqrt{1+F_{23}^2+F_{34}^2}}=2(ab-cd),\\
\displaystyle \frac{F_{23}}{\sqrt{1+F_{23}^2+F_{34}^2}}=2(bc+ad),
\end{array}
\label{eq:tduality3D}
\end{eqnarray}
together with the normalization condition $a^2+b^2+c^2+d^2=1$ and the constraint $F_{24}=0$.\\\\
Suppose we want to get a brane stretching along the diagonal of a 3-cube. The rotation
which rotates the $x_2$ axis into the diagonal has the rotation axis $(b,0,b)^T$. For the rotation
angle holds $\mbox{arccos}\left(\theta\right)=1/\sqrt{3}$. The quaternion describing this rotation is,
\begin{eqnarray}
R_L=\mbox{arccos}(\frac{\theta}{2}) + b i + 0 j + b k.
\end{eqnarray}
The normalization condition fixes the value of the unknown at,
\begin{eqnarray}
b=\frac{1}{2}\sqrt{1-\frac{1}{\sqrt{3}}}.
\end{eqnarray}
From the T-duality correspondence Eq.~(\ref{eq:tduality3D}) we see that such a D-brane stretching
across the diagonal of a cube on the IIA side is T-dual to a brane with the magnetic fluxes,
\begin{eqnarray}
F_{23}=F_{34}=1,
\end{eqnarray} 
and all other components zero. The supersymmetry condition Eq.~(\ref{eq:susycond}) is however not satisfied for this
example.
\subsection{T-Duality in 4D}
After the three-dimensional warm-up, let us now turn to the four-dimensional case in full generality.
By comparing the Cayley transform of the field strength tensor to Eq.~(\ref{eq:Cayley1}) we obtain the correspondence,
\begin{eqnarray}
\begin{array}{rcl}
\displaystyle a^2-b^2+c^2-d^2&=&\displaystyle \frac{1}{N}(1-F_{14} F_{23}-F_{12} F_{34})\\
\displaystyle  -2 (a b - c d)&=&\displaystyle -\frac{1}{N}(F_{12}+F_{34})\\
\displaystyle  0&=&0 \\
\displaystyle  -2(b c + a d)&=& \displaystyle -\frac{1}{N}(F_{14}+F_{23})\\
\\
\displaystyle p^2-q^2+r^2-s^2&=&\displaystyle \frac{1}{N}(1+F_{14} F_{23}+F_{12} F_{34})\\
\displaystyle  -2 (p q + r s)&=&\displaystyle -\frac{1}{N}(F_{12}-F_{34})\\
\displaystyle  0&=&0 \\
\displaystyle   2 (q r - p s)&=&\displaystyle -\frac{1}{N} (F_{14}-F_{23}),
\end{array}
\end{eqnarray}
with,
\begin{eqnarray}
N=\sqrt{{(F_{14}+F_{23})}^2+\left(F_{12}+F_{34}\right)^2 + \left(1-F_{14} F_{23}-F_{12} F_{34}\right)^2}.
\end{eqnarray}
In order to solve the system of equations, the normalization condition Eq.~(\ref{eq:quatnorm}) for quaternions has to be imposed as well. In addition one must also require that the rotation reduces to the identity in the limit of vanishing field strenght tensor. The solution for $R_L$ and $R_R$ is,
\begin{eqnarray}
\begin{array}{rclrcl}
a&=&N_{-} & p&=&N_{+},\\
b&=&\displaystyle \frac{1}{2 N N_{-}}(F_{12}+F_{34}) & q&=&\displaystyle \frac{1}{2 N N_{+}}(F_{12}-F_{34}),\\
c&=&0 & r&=&0,\\
d&=&\displaystyle \frac{1}{2 N N_{-}}(F_{14}+F_{23}) & s&=&\displaystyle \frac{1}{2 N N_{+}}(F_{14}-F_{23}),
\end{array}
\end{eqnarray}
with,
\begin{eqnarray}
N_{\pm}=\sqrt{\frac{1}{2}+\frac{1}{2 N}(1\pm F_{14} F_{23} \pm F_{12} F_{34})}.
\end{eqnarray}
The world-sheet of a D2-brane dual to an unmagnetized D4-brane is spanned by the vectors $u=(1,0,0,0)$ and $v=(0,0,1,0)$. The world-sheet of a D2-brane dual to a D4-brane with arbitrary fluxes $|F_{ij}|<\infty$ is spanned by the vectors,
\begin{eqnarray}
u'=R_L u R_R\qquad v'=R_L v R_R.\label{eq:rotmat}
\end{eqnarray}
For completeness note that the brane on the IIA side may also be a coisotropic D4-brane; I do not pursued this further here.\\
When all fluxes are vanishing on the IIB side, the homology class of the IIA brane is $dx^1 \wedge dx^2 \wedge dx^3$.
Applying this rotation to the differentials of the above homology class, the homology class of the magnetized 2-cycle is obtained:
\begin{eqnarray}
\begin{array}{rcl}
[\Pi]&=&dx^1\wedge dx^2 - F_{23}\; dx^1 \wedge dy^1 + F_{34}\; dx^1 \wedge dy^2
+ F_{12}\; dy^1\wedge dx^2 \\
&&-F_{14}\; dx^2 \wedge dy^2 + (F_{14} F_{23}+F_{12} F_{34})\; dy^1 \wedge dy^2. 
\end{array}
\end{eqnarray}
\subsection{Example}
From Eq.~(\ref{eq:rotmat}) it can be worked out how the brane stretching for example along the diagonal of a 3D cube can be constructed. With one coordinate in 4D fixed, the cube located at that point should contain a brane that stretches diagonally along its volume. Clearly, the point in 4D must be one of the orbifold fixed points in order to peg the brane to it.\\
For this to happen at least at one fixed point, at least three of the four fields $F_{12}, F_{14}, F_{23}$ and $F_{34}$
must be non-vanishing.
A very symmetric choice turns out to be the one when all four fields are set to $F_{12}=1$, $F_{14}=1$, $F_{23}=-1$ and $F_{34}=-1$. The A-side brane is parametrized by,
\begin{eqnarray}
M: \begin{pmatrix}u \\ 0 \\ v \\ 0\end{pmatrix}
\mapsto \frac{1}{\sqrt{3}}
\begin{pmatrix}u \\ u+v \\ v \\ 
u-v \end{pmatrix}=
u \begin{pmatrix}1 \\ 1 \\ 0 \\ 1\end{pmatrix}
+v \begin{pmatrix}0 \\ 1 \\ 1 \\ -1\end{pmatrix}.\label{eq:tentbrane}
\end{eqnarray}
Of course a constant vector is to be added when the position moduli assume non-vanishing values.
The homology class of this brane is,
\begin{eqnarray}
\begin{array}{rcl}
[\Pi_M]&=&dx^1\wedge dx^2 +  dx^1 \wedge dy^1 -  dx^1 \wedge dy^2
+ dy^1\wedge dx^2 -dx^2 \wedge dy^2 - 2\; dy^1 \wedge dy^2. \label{eq:homclass}
\end{array}
\end{eqnarray}
\section{Supersymmetry or the Absence of Tachyons}
\label{sec:susy}
\subsection{IIA side}
Phenomenologically interesting brane-worlds typically exhibit $\mathcal{N}=1$ supersymmetry.
In such a setting, two branes D$6_a$ and D$6_b$ on the IIA side are related by a $SU(3)$-rotation
The orientation of the matrix determines the chirality of the fermion at the intersection. Similarly, every brane is related by an $SU(3)$-rotation with the O-plane. The eigenvalues have an interpretation of rotation angles and give the spectrum of light scalars. Here, we are only interested in the observable low-energy physics so the light scalars will not concern us unless their uncorrected mass is exactly zero.\\
In the framework used in this paper there would be no need to refer to the $SU(3)$-rotation to test for supersymmetry since the LG model is supersymmetric by construction. The equivalent condition would be the absence of tachyons which is ensured by selecting branes of appropriate R-charges as was discussed in~\cite{harunT6}. It is useful to discuss the $SU(3)$ condition nevertheless -- it allows a convenient preselection of the homology classes we are interested in, before actually constructing their factorization which is a very tedious task. Moreover, nearly all work in the field has been done on factorizable cycles so it is worthwhile to discuss the $SU(3)$ rotation for once for more general cycles. For factorizable cycles, the rotation amounts merely to planar rotations $\phi_{1,2,3}$ on each of the three two-tori $1,2,3$. The rotation is $SU(3)$ if the sum of rotation angles vanishes $\phi_1+\phi_2+\phi_3=0$. For non-factorizable cycles, we need to look at the actual $SU(3)$-rotation matrix. It could of course be generated by
the Gell-Mann matrices, but a more convenient parametrization is advantageous. Indeed there is a very simple form, provided that at least one matrix entry is vanishing. It looks as follows:
\begin{eqnarray}
R(r_i,s_i)=\begin{pmatrix}
\displaystyle\frac{r_1+i s_1}{N_1} &  \displaystyle  \frac{r_2-i s_2}{N_1} \frac{r_4-i s_4}{N_2} & \displaystyle  \frac{r_2-i s_2}{N_1} \frac{r_3-i s_3}{N_2}\\
\displaystyle \frac{r_2+i s_2}{N_1} &  \displaystyle  \frac{r_1-i s_1}{N_1} \frac{r_4-i s_4}{N_2} & \displaystyle  \frac{r_1-i s_1}{N_1} \frac{r_3-i s_3}{N_2}\\
0 & \displaystyle  \frac{r_3+i s_3}{N_2} & \displaystyle  \frac{r_4+i s_4}{N_2}
\end{pmatrix},
\end{eqnarray}
with $N_1=\sqrt{r_1^2+s_1^2+r_2^2+s_2^2}$ and $N_2=\sqrt{r_3^2+s_3^2+r_4^2+s_4^2}$. For this matrix $R R^{\dagger}=\mathbbm{1}$ holds for any real values of $r_i$ and $s_j$.\\\\
Take a generic two cycle that wraps one of the fundamental cycles on the third two-torus of an $\textbf{AAA}$-lattice for example. 
As a 3-plane in flat 6D space the locus of the orientifold plane in the $[a]\times [a] \times [a]$ homology class is decribed by the set $\left\{(u,0,v,0,w,0) | u,v,w\in \mathbb{R}^6 \right\}$. In complex notation the basis is spanned by the three unit vectors
$(1,0,0),(0,1,0),(0,0,1) \in \mathbb{C}^3$. Their images under the $SU(3)$ rotation, that is the column vectors of the matrix, span the
3-plane of the rotated brane. A rotation into a 2-cycle $\times$ fundamental 1-cycle is therefore given by,
\begin{eqnarray}
\begin{array}{l}
R(c_1,d_1,c_2,d_2;0)= \displaystyle\frac{1}{\sqrt{r_1^2+s_1^2+r_2^2+s_2^2}}
\begin{pmatrix}
r_1 +i s_1 & r_2 +i s_2\\
-r_2 +i s_2 & r_1 -i s_1
\end{pmatrix}
\oplus
\begin{pmatrix}
1
\end{pmatrix},
\\\\
R(r_1,s_1,r_2,s_2;\frac{1}{2}\pi)=\displaystyle \frac{1}{\sqrt{r_1^2+s_1^2+r_2^2+s_2^2}}
\begin{pmatrix}
r_1 +i s_1 & s_2 -i r_2\\
-r_2 +i s_2 & -r_1 -i s_1
\end{pmatrix}
\oplus
\begin{pmatrix}
i
\end{pmatrix}.
\end{array}
\end{eqnarray}
If we chose the two-cycle to be the one from Eq.~(\ref{eq:tentbrane}) we find that,
\begin{eqnarray}
\begin{array}{l}
R(r_1=1,s_2=1,r_2=0,s_2=1;0)=\displaystyle \frac{1}{\sqrt{3}}
\begin{pmatrix}
1 +i & i\\
i  & 1 -i
\end{pmatrix}
\oplus
\begin{pmatrix}
1
\end{pmatrix}
\end{array}
\end{eqnarray}
contains the desired 2-cycle: The first column of the matrix corresponds to the vector $(1,1,0,1)^T\in \mathbb{R}^4$ of Eq.~(\ref{eq:tentbrane}) and
the second column to $(0,1,1,-1)^T\in \mathbb{R}^4$. Supersymmetry is therefore preserved by that cycle. If we want it to wrap the other fundamental cycle on the third two-torus, a $SU(3)$ rotation can not be found, therefore supersymmetry is broken.\\\\
What is the homology class of a brane given by such a rotation? 
The orientifold plane's homology charge $[\Pi_{O6}]=dx^1\wedge dx^2 \wedge dx^3$ should be related by the same transformation to the
homology charge of the resulting brane. Up to a prefactor we get:
\begin{eqnarray}
\begin{array}{l}
x^1 \rightarrow x^1 + y^1 + y^2\\
x^2 \rightarrow y^1 + x^2 - y^2\\
x^3 \rightarrow x^3
\end{array}
\end{eqnarray}
The homology class of the $D6$ brane is of course $[\Pi_M]$ again, this time times the wedge of $dx^3$.\\\\
\subsection{IIB side}
Provided that the field strength tensor $F$ is known, the computation can of course be done on the B-side~\cite{MMWS}.
It is straightforward but for completeness it is done here.
In order to obtaine a consistent $\mathcal{N}=1$ SUSY theory in 4d, two constraints apart from tadpole cancellation have to be satisfied for a magnetized D9-brane on the IIB side~\cite{MMWS}:
\begin{eqnarray}
\mbox{tan}\; \theta (\frac{1}{2!}J\wedge J \wedge \mathcal{F}^a- \frac{1}{3!}\mathcal{F}^a\wedge \mathcal{F}^a \wedge \mathcal{F}^a)
&=&\frac{1}{3!}J\wedge J \wedge J - \frac{1}{2!}J \wedge \mathcal{F}^a \wedge \mathcal{F}^a.\\
\mathcal{F}^a_{(2,0)}&=&0,\label{eq:susycond2}
\end{eqnarray} 
Here $\theta$ is the phase of the branes. Rewriting the field strength in complex coordinates,
\begin{eqnarray}
\mathcal{F}^a=-2\pi i \alpha'
\begin{pmatrix}
F^a_{(2,0)} & F^a_{(1,1)} \\
-F^{a\dagger}_{(1,1)} & F^{a*}_{(2,0)} 
\end{pmatrix},
\end{eqnarray} 
it decomposes into the two 3x3 matrices (see e.g.~\cite{antoniadismaillard}),
\begin{eqnarray}
F^a_{(2,0)}&=&2\pi \left.(\tau-\bar \tau)^{-1}\right.^T\left[\tau^T p^a_{xx}\tau -\tau^T p^a_{xy}-p^a_{yx}\tau +p^a_{yy}\right](\tau-\bar \tau)^{-1},\\
F^a_{(1,1)}&=&2\pi \left.(\tau-\bar \tau)^{-1}\right.^T\left[-\tau^T p^a_{xx}\bar \tau +\tau^T p^a_{xy}+p^a_{yx}\bar \tau -p^a_{yy}\right](\tau-\bar \tau)^{-1}.
\end{eqnarray} 
The matrices $(p^a_{xx})_{ij}$, $(p^a_{xy})_{ij}$ and $(p^a_{yy})_{ij}$ come from the field strength in the directions
$(x^i,x^j)$, $(x^i,y^j)$ and $(y^i,y^j)$. By construction holds,
\begin{eqnarray}
(p^a_{xy})_{ij} dx^i \wedge dy^j=-(p^a_{xy})_{ij} dy^j \wedge dx^i = (p^a_{yx})_{ji} dy^j \wedge dx^i,
\end{eqnarray}
and therefore $p^a_{xy}=-(p^a_{yx})^T$.
The corresponding Dirac quantization condition is,
\begin{eqnarray}
q_a F^a_{kl}=2\pi \frac{m^a_{kl}}{n^a_{kl}}\equiv 2\pi p^a_{kl}.
\end{eqnarray}
Using the above decomposition, Eq.~(\ref{eq:susycond2}) becomes,
\begin{eqnarray}
\tau^T p^a_{xx}\tau -\tau^T p^a_{xy}-p^a_{yx}\tau +p^a_{yy}=0.
\end{eqnarray}
In a $\mathbbm{T}^6$-orbifold, all off-diagonal elements of $\tau$ are projected out. In the case of square tori, i.e. $\tau_{11}=\tau_{22}=\tau_{33}=i$, the real and imaginary part of this equation reduce to,
\begin{eqnarray}
p^a_{xx}=p^a_{yy} \qquad \mbox{and}\qquad p^a_{xy}=(p^a_{xy})^T. \label{eq:squaretori}
\end{eqnarray}
Let us write out these matrices explicitly:
\begin{eqnarray}
p_{xx}=
\begin{pmatrix}
0 & F_{13} & F_{15} \\ -F_{13} & 0 & F_{35}\\ -F_{15} & -F_{35} & 0
\end{pmatrix}
&
p_{yy}=
\begin{pmatrix}
0 & F_{24} & F_{26} \\ -F_{24} & 0 & F_{46}\\ -F_{26} & -F_{46} & 0
\end{pmatrix}\\
p_{xy}=
\begin{pmatrix}
F_{12} & F_{14} & F_{16} \\ -F_{23} & F_{34} & F_{36}\\ -F_{25} & -F_{45} & F_{56}
\end{pmatrix}
\end{eqnarray}
The $b_2=15$ independent components of the field strength tensor $F$, reduce to 9 after imposing the conditions of Eq.~(\ref{eq:squaretori}):
\begin{eqnarray}
\begin{array}{rclrclrcl}
\displaystyle F_{24}&=&F_{13},  & F_{26}&=&F_{15}, & F_{46}&=&F_{35},\\
\displaystyle -F_{23}&=&F_{14}, & -F_{25}&=&F_{16}, & -F_{45}&=&F_{36}.
\end{array} \label{eq:susycond}
\end{eqnarray}
The field strenght tensor becomes:
\begin{eqnarray}
F=
\begin{pmatrix}
     0  & F_{12} & F_{13} & F_{14} & F_{15} & F_{16}\\
-F_{12} & 0      &-F_{14} & F_{13} &-F_{16} & F_{15}\\
-F_{13} & F_{14} & 0      & F_{34} & F_{35} & F_{36}\\
-F_{14} &-F_{13} &-F_{34} & 0      &-F_{36} & F_{35}\\
-F_{15} & F_{16} &-F_{35} & F_{36} & 0      & F_{56}\\
-F_{16} &-F_{15} &-F_{36} &-F_{35} &-F_{56} & 0     \\
\end{pmatrix}.
\end{eqnarray}
\subsection{Matrix Factorizations}
When dealing with matrix factorizations, one can extract their phase
and compare it with the phase of  the orientifold plane.
The phase of the branes also determines the charges of the morphisms.
The issue of supersymmetry and the absence of tachyons has already been addressed in~\cite{harunT6}.
\section{Yukawa Couplings}
We are now in a position to set up D-branes, find the open strings stretching between
them and compute Yukawa couplings. From section~\ref{sec:loc} we know how to relate
the physical locus of the IIA brane to its homology class and from section~\ref{sec:susy}
we know how to test for supersymmetry. Given a satisfactory set of branes we can finally
construct a matrix factorization for a brane in the desired homology class and then shift it
to any position or leave the position modulus as a parameter in the theory. The next step
is to find the open string states and compute the Yukawa couplings from them.
It would have been desirable to present a complete model here with all its Yukawa couplings.
But given that a typical matrix requires about half a page to write down, it would
be pointless to add dozens of pages with matrices here, as long as the model does not possess
any unusual features providing new insights. For demonstration purposes it should be sufficient
to compute a few Yukawa couplings as examples. Three non-factorizable
branes -- $Q_7$, $Q_8$ and $Q_9$ -- which can give rise to intersection numbers of three
are listed in Appendix~\ref{sec:morphlist} together with three open string states.
For the reader who has skipped parts of this paper, let me summarize briefly to what extent
the computation is generic. No attempt has been made to select particularly simple branes.
Branes with any homology class can be constructed easily. From the morphisms between the branes
it is also easy to compute the Yukawa couplings and here, too, nothing was tuned to be simple.
There is only one restriction to full generality, namely the brane moduli. The branes considered
here pass through the fixed points of all three two-tori. In section~\ref{sec:moduli} it was
explained how the branes can be moved away from the fixed points in one two-torus or
even moved completely into the bulk by letting the moduli assume generic values everywhere.
In these settings, however, it is significantly more cumbersome to find a basis for the cohomology. 
Fortunately, the latter case is phenomenologically of little interest for the orbifolds discussed here.
The multiplicity of intersections of bulk cycles are multiples of the rank of the orbifold gauge group. An odd number
of families can therefore not be achieved. This would be different for $\mathbb{Z}_3$-orbifolds, of course.\\\\
The three-point correlators are determined by a residue integral, which for $n$ Landau-Ginzburg fields reads~\cite{kapustinli}:
\begin{eqnarray}
\langle \Psi_a \Psi_b \Psi_c \rangle =\oint{\frac{1}{n!}\frac{\mbox{STr}\;Q^{\wedge n}\Psi_a \Psi_b \Psi_c}{\partial_1 W...\partial_n W}}.\label{eq:kapli}
\end{eqnarray}
The correlator which can be computed from the morphism in Appendix~\ref{sec:morphlist} is:
\begin{eqnarray}
\langle \Psi^{(7\rightarrow 8)} \Psi^{(8\rightarrow 9)} \Psi^{(9\rightarrow 7)}\rangle
=\displaystyle\frac{c(\tau_1)^4 c(\tau_2) (-1+c(\tau_3))^2 (1+c(\tau_3))^3}{\left(-1+c(\tau_1)^4\right)^2 \left(-1+c(\tau_2)^2\right)c(\tau_3)^2}.
\end{eqnarray}
The morphisms and the correlator chosen here have been selected only for illustration, not for any particular physical significance.
Some further correlators take the following form:
\begin{eqnarray}
\begin{array}{l}
Y_1=\displaystyle\frac{c(\tau_1)^2 (1+c(\tau_1)) c(\tau_2)}{\left(1+c(\tau_1)^2\right)^2 \left(-1+c(\tau_2)^2\right)}\\
\\
Y_2=\displaystyle\frac{c(\tau_1)^4 (-1+c(\tau_2)) c(\tau_2)^2 (-1+c(\tau_3))^2 (1+c(\tau_3))^3}{8 (1+c(\tau_1)) \left(-1+c(\tau_1)-c(\tau_1)^2+c(\tau_1)^3\right)^2
\left(1+c(\tau_2)^2\right)^2 c(\tau_3)^2}\\
\\
Y_3=\displaystyle\frac{c(\tau_1)^3 c(\tau_2) \left(1-c(\tau_2)+c(\tau_2)^2\right)}{2 \left(1+c(\tau_1)^2\right)^2 (-1+c(\tau_2))
\left(1+c(\tau_2)^2\right)^2}\\
\\
Y_4=\displaystyle\frac{c(\tau_1)^2 c(\tau_2) \left(1-c(\tau_2)+c(\tau_2)^2\right) (1+c(\tau_3)) \left(c(\tau_3)^2+c(\tau_1)^4 c(\tau_3)^2-c(\tau_1)^2 \left(-1+c(\tau_3)^2\right)^2\right)}{4(-1+c(\tau_1)) \left(1+c(\tau_1)+c(\tau_1)^2+c(\tau_1)^3\right)^2 (-1+c(\tau_2)) \left(1+c(\tau_2)^2\right)^2 c(\tau_3)^2}
\end{array}
\end{eqnarray}
Again, these correlators should only be regarded as representatives illustrating the feasibility of the method. For that reason I refrain from listing
the large matrices used in the computation. As usual, in these results $\tau_{1,2,3}$ are the complex (K\"ahler) structure of the three two-tori on the B-side (A-side).
The functions $c_i \equiv c(\tau_i)$ were given in Eq.~(\ref{eq:ci}) in terms of Jacobi theta functions.
The morphisms which were plugged into Eq.~(\ref{eq:kapli}) to obtain the correlators $Y_j$ are not all given here.
\appendix
\section{Transformation to the Bundle Data and Wrapping Numbers}
In order to obtain the bundle data and wrapping number for branes on the $\mathbbm{T}^4$ and $\mathbbm{T}^6$, we need to tensor the results from the $\mathbbm{T}^2$. A D2 brane on the first torus tensored with a bound state of a D2 brane and an D0-antibrane on the second one for instance is a D4-brane with a D2-antibrane on the first torus. The analog to Eq.~(\ref{eq:chern}) together with the factor of two from the hyperplane intersections therefore gives for the $\mathbbm{T}^4$ and the $\mathbbm{T}^6$,
\begin{eqnarray}
\mbox{ch}(Q_{tensor}^{2\otimes})=
\begin{pmatrix}
1 & 1 & 1 & 1\\
0 & -1 & 0 & -1\\
0 & 0 & -1 & -1\\
0 & 0 & 0 & 1
\end{pmatrix},
\end{eqnarray}
\begin{eqnarray}
\mbox{ch}(Q_{tensor}^{3\otimes})=
\begin{pmatrix}
1 & 1 & 1 & 1 & 1 & 1 & 1 & 1\\
0 & -2 & 0 & -2 & 0 & -2 & 0 & -2\\
0 & 0 & -2 & -2 & 0 & 0 & -2 & -2\\
0 & 0 & 0 & 4 & 0 & 0 & 0 & 4\\
0 & 0 & 0 & 0 & -2 &-2 &-2 &-2\\
0 & 0 & 0 & 0 & 0 & 4 & 0 & 4\\
0 & 0 & 0 & 0 & 0 & 0 & 4 & 4\\
0 & 0 & 0 & 0 & 0 & 0 & 0 & 8
\end{pmatrix}.
\end{eqnarray}
The entries of the charge matrix are $\langle Q^{(n_1,n_2)} | k_1,k_2 \rangle$ and $\langle Q^{(n_1,n_2,n_3)} | k_1,k_2,k_3 \rangle$ respectively, which we sort according to Eq.~(\ref{eq:groupordering}).
The transformations to large-volume then read,
\begin{eqnarray}
T^{LV}_{T^4}=
\frac{1}{64}\begin{pmatrix}
2i & 2 & 2 & -2i\\
-2-2i & -2-2i & -2+2i & -2+2i\\
-2-2i & -2+2i & -2-2i & -2+2i\\
4 & 4 & 4 & 4
\end{pmatrix},
\end{eqnarray}
and,
\begin{eqnarray}
T^{LV}_{T^6}=
\frac{1}{256}\begin{pmatrix}
1-i & -1-i & -1-i & -1+i & -1-i & -1+i & -1+i & 1+i\\
2i & 2i & 2 & 2 & 2 & 2 & -2i & -2i\\
2i & 2 & 2i & 2 & 2 & -2i & 2 & -2i\\
-2-2i & -2-2i & -2-2i & -2-2i & -2+2i & -2+2i & -2+2i & -2+2i\\
2i & 2 & 2 & -2i & 2i & 2 & 2 & -2i\\
-2-2i & -2-2i & -2+2i & -2+2i & -2-2i & -2-2i & -2+2i & -2+2i\\
-2-2i & -2+2i & -2-2i & -2+2i & -2-2i & -2+2i & -2-2i & -2+2i\\
4 & 4 & 4 & 4 & 4 & 4 & 4 & 4
\end{pmatrix}.\nonumber
\end{eqnarray}
Finally, in order to obtain the A-side charges, Eq.~(\ref{eq:Aside}) has to be applied likewise, which gives the transformation to the A-side wrapping numbers:
\begin{eqnarray}
T^{A}_{T^4}=
\begin{pmatrix}
1 & 0 & 0 & 0\\
1 & 1 & 0 & 0\\
1 & 0 & 1 & 0\\
1 & 1 & 1 & 1
\end{pmatrix},
\end{eqnarray}
and
\begin{eqnarray}
T^{A}_{T^6}=
\begin{pmatrix}
1 & 0 & 0 & 0 & 0 & 0 & 0 & 0\\
1 & 1 & 0 & 0 & 0 & 0 & 0 & 0\\
1 & 0 & 1 & 0 & 0 & 0 & 0 & 0\\
1 & 1 & 1 & 1 & 0 & 0 & 0 & 0\\
1 & 0 & 0 & 0 & 1 & 0 & 0 & 0\\
1 & 1 & 0 & 0 & 1 & 1 & 0 & 0\\
1 & 0 & 1 & 0 & 1 & 0 & 1 & 0\\
1 & 1 & 1 & 1 & 1 & 1 & 1 & 1
\end{pmatrix}.
\end{eqnarray}
\section{List of Branes and Open String States}
\label{sec:morphlist}
One particular factorization on the $\mathbb{T}^6$-orbifolds is shown here:
\begin{eqnarray}
\begin{array}{l}
E_7=\begin{pmatrix}
X_1 & Y_1 & 0 & -Y_3\\
Y_2 Y_3 Y_4 & -X_2 X_3 X_4 & -X_2 Y_2 Y_3 \\
0 & 0 & -Y_1 Y_2 & -X_3 X_4 \\
0 & 0 & X_1 X_2 & -Y_3 Y_4
\end{pmatrix}\otimes \begin{pmatrix} Z_1 \end{pmatrix} \otimes \begin{pmatrix} z_1 \end{pmatrix},\\\\
J_7=\begin{pmatrix}
X_2 X_3 X_4 & Y_1 & -X_2 Y_3 & 0\\
Y_2 Y_3 Y_4 & -X_1 & 0 & -Y_2 Y_3 \\
0 & 0 & -Y_3 Y_3 & X_3 X_4 \\
0 & 0 & -X_1 X_2 & -Y_1 Y_2
\end{pmatrix}\otimes \begin{pmatrix} -Z_2 Z_3 Z_4 \end{pmatrix} \otimes \begin{pmatrix} -z_1 \end{pmatrix}.
\end{array}
\label{eq:Q6}
\end{eqnarray}
The brane has been obtained by a series of tachyon condensation processes according to Eqs.~(\ref{eq:tachcon1})-(\ref{eq:tachcon2}). 
The building blocks of the factorizations $Q_a$ and $Q_b$ can still be recognized in Eq.~(\ref{eq:Q6}).\\
The variables $X,Y,Z$ are associated with the first, second and third torus respectively. 
By permuting tori, other factorizations can be obtained easily. We define $Q_8$ by taking $Q_7$ and replacing,
\begin{eqnarray}
(X_i,Y_j,Z_k) \rightarrow (Z_i,Y_j,X_k) \qquad i,j,k=\{1,2,3,4\},
\end{eqnarray} 
whereas in $Q_9$ we substitute,
\begin{eqnarray}
(X_i,Y_j,Z_k) \rightarrow (Z_i,X_j,Y_k) \qquad i,j,k=\{1,2,3,4\}.
\end{eqnarray} 
These branes are invariant under $g_0, g_1, g_2, g_3, g_4, h_1, h_2$, that is under all orbifold actions considered here.
Their orbifold generators are,
\begin{eqnarray}
\begin{array}{rcl@{}r@{}r@{}r@{}r@{}r@{}r@{}r@{}r@{}r@{}r@{}r@{}r@{}r@{}r@{}r@{}r@{}r}
\gamma^7(g_0)&=&\mbox{diag }(&1,& -1,& -i,& -i,& -1,& -1,& -1,& -1,& i,& i,& i, &i,& i,& -i,& 1,& \;1),\\
\gamma^7(g_1)&=&\mbox{diag }(&1,& i,&-1,& 1,& i,& 1,& -1,& 1,& i,& 1,& -1,& 1,& 1,& i,& -1,& \;1),\\
\gamma^7(g_2)&=&\mbox{diag }(&i,& -1,& -1,& 1,& i,& -1,& 1,& -1,& i,& -1,& 1,& -1,& i,& -1,& -1,& \;1),\\
\gamma^7(g_3)&=&\mbox{diag }(&-i,& -i,& -i,& -i,& 1,& 1,& 1,& 1,& -i,& -i,& -i,& -i,& 1,& 1,& 1,& \;1),\\
\gamma^7(h_1)&=&\mbox{diag }(&1, &-1,& 1,& 1,& -1,& 1,& 1,& 1,& -1,& 1,& 1,& 1,& 1,& -1,& 1,&\; 1),\\
\gamma^7(h_2)&=&\mbox{diag }(&1, &1,& 1,& 1,& 1,& 1,& 1,& 1,& 1,& 1,& 1,& 1,& 1,& 1,& 1,& \;1),
\end{array}
\end{eqnarray}
\begin{eqnarray}
\begin{array}{rcl@{}r@{}r@{}r@{}r@{}r@{}r@{}r@{}r@{}r@{}r@{}r@{}r@{}r@{}r@{}r@{}r@{}r}
\gamma^8(g_0)&=&\mbox{diag }(&1,& -1,& -i,& -i,& -1,& -1,& -1,& -1,& i,& i,& i,& i,& i,& -i,& 1,& 1),\\
\gamma^8(g_1)&=&\mbox{diag }(&-i,& -i,& -i,& -i,& 1,& 1,& 1,& 1,& -i,& -i,& -i,& -i,& 1,& 1,& 1,& 1),\\
\gamma^8(g_2)&=&\mbox{diag }(&i,& -1,& -1,& 1,& i,& -1,& 1,& -1,& i,& -1,& 1,& -1,& i,& -1,& -1,& 1),\\
\gamma^8(g_3)&=&\mbox{diag }(&1,& i,& -1,& 1,& i,& 1,& -1,& 1,& i,& 1,& -1,& 1,& 1,& i,& -1,& 1),&\\
\gamma^8(h_1)&=&\mbox{diag }(&-1,& -1,& -1,& -1,& 1,& 1,& 1,& 1,& -1,& -1,& -1,& -1,& 1,& 1,& 1,& 1),\\
\gamma^8(h_2)&=&\mbox{diag }(&1,& 1,& 1,& 1,& 1,& 1,& 1,& 1,& 1,& 1,& 1,& 1,& 1,& 1,& 1,& 1),
\end{array}
\end{eqnarray}
\begin{eqnarray}
\begin{array}{rcl@{}r@{}r@{}r@{}r@{}r@{}r@{}r@{}r@{}r@{}r@{}r@{}r@{}r@{}r@{}r@{}r@{}r}
\gamma^9(g_0)&=&\mbox{diag }(&1,& -1,& -i,& -i,& -1,& -1,& -1,& -1,& i,& i,& i,& i,& i,& -i,& 1,& 1),\\
\gamma^9(g_1)&=&\mbox{diag }(&i,& -1,& -1,& 1,& i,& -1,& 1,& -1,& i,& -1,& 1,& -1,& i,& -1,& -1,& 1),\\
\gamma^9(g_2)&=&\mbox{diag }(&-i,& -i,& -i,& -i,& 1,& 1,& 1,& 1,& -i,& -i,& -i,& -i,& 1,& 1,& 1,& 1),\\
\gamma^9(g_3)&=&\mbox{diag }(&1,& i,& -1,& 1,& i,& 1,& -1,& 1,& i,& 1,& -1,& 1,& 1,& i,& -1,& 1),\\
\gamma^9(h_1)&=&\mbox{diag }(&-1,& 1,& 1,& 1,& -1,& 1,& 1,& 1,& -1,& 1,& 1,& 1,& -1,& 1,& 1,& 1),\\
\gamma^9(h_2)&=&\mbox{diag }(&1,& 1,& -1,& 1,& 1,& 1,& 1,& -1,& 1,& 1,& 1,& -1,& 1,& 1,& -1,& 1).
\end{array}
\end{eqnarray}
The listed generators $g_j$ can be multiplied by $i^n,\;\;n\in \mathbb{N}$ and the generators
$g_j$ can be multiplied by $(-1)^n,\;\;n\in \mathbb{N}$ to get the other orbits of the branes.
The offset of the phase has been chosen arbitrarily here so that only phase differences matter.
According to Eqs.~(\ref{eq:homT6a})-(\ref{eq:homT6b}) the homology class of the
branes $(Q_k,\gamma^{k}(g_1),\gamma^{k}(g_2),\gamma^{k}(g_3))$, $k=7,8,9$ is given by,
\begin{eqnarray}
\vec{q}_7=(0,0,0,0,-1,2,1,-1),\\
\vec{q}_8=(0,-1,0,2,0,1,0,-1),\\
\vec{q}_9=(0,0,-1,2,0,0,1,-1).
\end{eqnarray}
A number of open string states exists between the branes. Three of them are listed in the following.
All of them are fermions so their block-structure is,
\begin{eqnarray}
\Psi^{(i\rightarrow j)}=
\begin{pmatrix}
0 & \psi^{(i\rightarrow j)}_0\\
\psi^{(i\rightarrow j)}_1 & 0\\
\end{pmatrix}.
\end{eqnarray}
The block matrices are:
\begin{eqnarray}
\begin{array}{l}
\psi^{(7\rightarrow 8)}_0=\begin{pmatrix}
0 & 0 & 0 & 1 & 0 & 0 & 0 & 0\\
0 & 0 & 0 & 0 & 0 & 0 & 0 & Y_2\\
0 & 0 & 0 & 0 & 0 & 0 & 0 & 0\\
0 & 0 & 0 & 0 & 0 & 0 & 0 & 0\\
0 & 0 & 0 & 0 & 0 & 0 & -Z_2 & 0\\
0 & 0 & Y_2 Z_2 & 0 & 0 & 0 & 0 & 0\\
0 & 0 & 0 & 0 & 0 & 0 & 0 & 0\\
0 & 0 & 0 & 0 & 0 & 0 & 0 & 0
\end{pmatrix}\otimes 
\begin{pmatrix} 1 \end{pmatrix},\\\\
\psi^{(7\rightarrow 8)}_1=\begin{pmatrix}
0 & 0 & 0 & 0 & 0 & 0 & 0 & 1\\
0 & 0 & 0 & Y_2 & 0 & 0 & 0 & 0\\
0 & 0 & 0 & 0 & 0 & 0 & 0 & 0\\
0 & 0 & 0 & 0 & 0 & 0 & 0 & 0\\
0 & 0 & Z_2 & 0 & 0 & 0 & 0 & 0\\
0 & 0 & 0 & 0 & 0 & 0 & -Y_2 Z_2 & 0\\
0 & 0 & 0 & 0 & 0 & 0 & 0 & 0\\
0 & 0 & 0 & 0 & 0 & 0 & 0 & 0\\
\end{pmatrix}\otimes 
\begin{pmatrix} 1 \end{pmatrix},
\end{array}
\end{eqnarray}
\begin{eqnarray}
\begin{array}{l}
\psi^{(8\rightarrow 9)}_0=\begin{pmatrix}
0 & 0 & 0 & 0 & 0 & 0 & 0 & 0\\
0 & 0 & 0 & 0 & -X_1 Y_2 Y_3 & 0 & 0 & Y_2 Y_3\\
0 & 0 & -Y_2 Z_1 & 0 & 0 & 0 & 0 & 0\\
0 & 0 & 0 & 0 & X_1 Z_1 & 0 & 0 & -Z_1\\
Y_2 Y_3 Z_1 & X_2 Y_2 Y_3 & 0 & 0 & 0 & 0 & 0 & 0\\
0 & 0 & 0 & 0 & 0 & 0 & 0 & 0\\
0 & 0 & 0 & 0 & Z_1 Z_3 Z_4 & 0 & -X_4 Z_1 & 0\\
Y_2 Z_1^2 & X_2 Y_2 Z_1 & 0 & 0 & 0 & 0 & 0 & 0
\end{pmatrix}\otimes 
\begin{pmatrix} 1 \end{pmatrix},\\\\
\psi^{(8\rightarrow 9)}_1=\begin{pmatrix}
X_1 Y_2 Y_3 & 0 & 0 & -Y_2 Y_3 & 0 & 0 & 0 & 0\\
0 & 0 & 0 & 0 & 0 & 0 & 0 & 0\\
0 & 0 & 0 & 0 & 0 & 0 & Z_1 & 0\\
X_1 Y_2 Z_1 & 0 & 0 & -Y_2 Z_1 & 0 & 0 & 0 & 0\\
0 & 0 & 0 & 0 & 0 & 0 & 0 & 0\\
0 & 0 & 0 & 0 & Y_2 Y_3 Z_1 & X_2 Y_2 Y_3 & 0 & 0\\
Y_2 Z_1 Z_3 Z_4 & 0 & -X_4 Y_2 Z_1 & 0 & 0 & 0 & 0 & 0\\
0 & 0 & 0 & 0 & -Z_1^2 & -X_2 Z_1 & 0 & 0
\end{pmatrix}\otimes 
\begin{pmatrix} 1 \end{pmatrix}
\end{array}
\end{eqnarray}
\begin{eqnarray}
\begin{array}{l}
\psi^{(9\rightarrow 7)}_0=\begin{pmatrix}
-Y_4 & 0 & X_2 & 0 & 0 & 0 & 0 & 0\\
0 & 0 & 0 & 0 & X_3 X_4 Y_4 & 0 & 0 & -X_3 X_4\\
0 & 0 & 0 & 0 & 0 & 0 & 0 & 0\\
0 & 0 & 0 & 0 & 0 & 0 & 0 & 0\\
0 & 0 & 0 & 0 & 0 & 0 & -Y_4 & 0\\
0 & X_3 X_4 Y_4 & 0 & 0 & 0 & 0 & 0 & 0\\
0 & 0 & 0 & 0 & 0 & 0 & 0 & 0\\
0 & 0 & 0 & 0 & 0 & 0 & 0 & 0\\
\end{pmatrix}\otimes 
\begin{pmatrix} 1 \end{pmatrix},\\\\
\psi^{(9\rightarrow 7)}_1=\begin{pmatrix}
0 & 0 & 0 & 0 & -Y_4 & 0 & X_2 & 0\\
X_3 X_4 Y_4 & 0 & 0 & -X_3 X_4 & 0 & 0 & 0 & 0\\
0 & 0 & 0 & 0 & 0 & 0 & 0 & 0\\
0 & 0 & 0 & 0 & 0 & 0 & 0 & 0\\
0 & Y_4 & 0 & 0 & 0 & 0 & 0 & 0\\
0 & 0 & 0 & 0 & 0 & -X_3 X_4 Y_4 & 0 & 0\\
0 & 0 & 0 & 0 & 0 & 0 & 0 & 0\\
0 & 0 & 0 & 0 & 0 & 0 & 0 & 0\\
\end{pmatrix}\otimes 
\begin{pmatrix} 1 \end{pmatrix}.
\end{array}
\end{eqnarray}
Remember that these morphisms exist between orbits of the branes with a certain phase difference $\Delta\phi$.
It is determined by demanding orbifold invariance:
\begin{eqnarray}
\gamma(g)\; \Psi(g x_i)\; (e^{\Delta\phi \pi i}\gamma(g))^{-1}=\Psi(x_i),
\end{eqnarray} 
where $g$ is the generator of the group.

\end{document}